\documentstyle[titlepage,12pt]{article}

\begin{document}

\renewcommand{\thesection}{\Roman{section}} \baselineskip=24pt plus 1pt
minus 1pt
\begin{titlepage}
\vspace*{0.5cm}

\begin{center}
\LARGE\bf 
Classical and Quantum Chaos in Atom Optics 
\\[1.5cm]
\normalsize\bf Farhan Saif
\end{center}

\vspace{7pt}
\begin{description}
\item Department of Electronics, Quaid-i-Azam University, Islamabad, Pakistan.
\item Department of Physics, University of Arizona, Tucson 85721, Arizona, USA.
\item saif@physics.arizona.edu, saif@fulbrightweb.org
\end{description}

\vspace{0.3cm}

\begin{center}
\normalsize 
The interaction of an atom with an electromagnetic field is
discussed in the presence of a time periodic external modulating
force. It is explained that a control on atom by electromagnetic
fields helps to design the quantum analog of classical optical
systems. In these atom optical systems chaos may appear at the onset
of external fields. The classical and quantum chaotic dynamics is
discussed, in particular in an atom optics Fermi accelerator. It is
found that the quantum dynamics exhibits dynamical localization and
quantum recurrences.
\end{center}


\end{titlepage}

\section{Introduction}

Two hundred years ago, G{\"o}ttingen physicist George Christoph
Lichtenberg wrote ``I think it is a sad situation in all our
chemistry that we are unable to suspend the constituents of matter
free''. Today, the possibilities to store atoms and to cool them to
temperatures as low as micro kelvin and nano kelvin scale, have made
atom optics a fascinating subject. It further provides a playground
to study the newer effects of quantum coherence and quantum
interference.

In atom optics we take into account the internal and external
degrees of freedom of an atom. The atom is considered as a de
Broglie matter wave and these are optical fields which provide
components, such as, mirrors, cavities and traps for the matter
waves (Meystre 2001, Dowling and Gea-Banacloche 1996). Thus, we find
a beautiful manifestation of quantum duality.

In atom optics systems another degree of freedom may be added by
providing an external periodic electromagnetic field. This
arrangement makes it feasible to open the discussion on chaos. These
periodically driven atom optics systems help to realize various
dynamical systems which earlier were of theoretical interest only.

The simplest periodically driven system, which has inspired the
scientists over many decades to understand various natural
phenomena, is Fermi accelerator. It is as simple as a ball bouncing
elastically on a vibrating horizontal plane. The system also
contributes enormously to the understanding of dynamical chaos
(Lichtenberg 1980, 1983, 1992).

\subsection{Atom optics: An overview}

In the last two decades, it has become feasible to perform
experiments with cold atoms in the realm of atom optics. Such
experiments have opened the way to find newer effects in the
behavior of atoms at very low temperature. The recent developments
in atom optics (Mlynek 1992, Adams 1994, Pillet 1994, Arimondo 1996,
Raizen 1999, Meystre 2001) make the subject a suitable framework to
realize dynamical systems.

Quantum duality, the work-horse of atom optics, is seen at work in
manipulating atoms. The atomic de Broglie waves are deflected,
focused and trapped. However, these are optical fields which provide
tools to manipulate the matter waves. As a consequence, in complete
analogy to classical optics, we have atom optical elements for
atoms, such as, mirrors, beam splitters, interferometers, lenses,
and waveguides (Sigel 1993, Adams 1994, Dowling 1996, Theuer 1999,
Meystre 2001).

As a manifestation of the wave-particle duality, we note that a
standing light wave provides optical crystal (Sleator 1992a, Sleator
1992b). Thus, we may find Raman-Nath scattering and Bragg scattering
of matter waves from an optical crystal (Saif 2001a, Khalique 2003).
In addition, an exponentially decaying electromagnetic field acts as
an atomic mirror (Balykin 1988, Kasevich 1990, Wallis 1992).

Atom interferometry is performed as an atomic de Broglie wave
scatters through two standing waves acting as optical crystal, and
aligned parallel to each other. The matter wave splits into coherent
beams which later recombine and create an atom interferometer (Rasel
1995). The atomic phase interferometry is performed as an atomic de
Broglie wave reflects back from two different positions of an atomic
mirror, recombines, and thus interferes (Henkel 1994, Steane 1995,
Szriftgiser 1996).

An atom optical mirror for atoms (Kazantsev 1974, Kazantsev 1990) is
achieved by the total internal reflection of laser light in a
dielectric slab (Balykin 1987, Balykin 1988). This creates, an
exponentially decaying electromagnetic field appears outside of the
dielectric surface. The decaying field provides an exponentially
increasing repulsive force to a blue detuned atom, which moves
towards the dielectric. The atom exhausts its kinetic energy against
the optical field and reflects back.

For an atom, which moves under gravity towards an atomic mirror, the
gravitational field and the atomic mirror together act like a
cavity---named as an atomic trampoline (Kasevich 1990) or a
gravitational cavity (Wallis 1992). The atom undergoes a bounded
motion in the system.

It is suggested by H. Wallis that a small change in the curvature of
the atomic mirror helps to make a simple surface trap for the
bouncing atom (Wallis 1992). An atomic mirror, comprising a blue
detuned and a red detuned optical field with different decay
constants, leads to the atomic trapping as well (Ovchinnikov 1991,
Desbiolles 1996).

The experimental observation of the trapping of atoms over an atomic
mirror in the presence of gravitational field was made by ENS group
in Paris (Aminoff 1993). In the experiment cold cesium atoms were
dropped from a magneto-optic trap on an atomic mirror, developed on
a concave spherical substrate, from a height of $3mm$. The bouncing
atoms were observed more than eight times, as shown in
figure~\ref{surfacetrap}.

\begin{figure}[t]
\begin{center}
\end{center}
\caption{Observation of trapping of atoms in a gravitational cavity:
(a) Schematic diagram of the experimental set up. (b) Number of
atoms detected by the probe beam after their initial release as a
function of time are shown as dots. The solid curve is a result of
corresponding Monte-Carlo simulations. (Aminoff 1993).}
\label{surfacetrap}
\end{figure}

A fascinating achievement of the gravitational cavity is the
development of recurrence tracking microscope (RTM) to study surface
structures with nano- and sub nano-meter resolutions (Saif 2001).
The microscope is based on the phenomena of quantum revivals.

In RTM, atoms observe successive reflections from the atomic mirror.
The mirror is joined to a cantilever which has its tip on the
surface under investigation. As the cantilever varies its position
following the surface structures, the atomic mirror changes its
position in the upward or downward direction. The time of a quantum
revival depends upon the initial height of the atoms above the
mirror which, thus, varies as the cantilever position changes.
Hence, the change in the time of revival reveals the surface
structures under investigation.

The gravitational cavity has been proposed (Ovchinnikov 1995,
S\"{o}ding 1995, Laryushin 1997, Ovchinnikov 1997) to cool atoms
down to the micro Kelvin temperature regime as well. Further cooling
of atoms has made it possible to obtain Bose-Einstein condensation
(Davis 1995, Anderson 1995, Bradley 1995), a few micrometers above
the evanescent wave atomic mirror (Hammes 2003, Rychtarik 2004).

A modulated gravitational cavity constitutes atom optics Fermi
accelerator for cold atoms(Chen 1997, Saif 1998). The system serves
as a suitable framework to analyze the classical dynamics and the
quantum dynamics in laboratory experiments. A bouncing atom displays
a rich dynamical behavior in the Fermi accelerator (Saif 1999).

\subsection{The Fermi accelerator}

In 1949, Enrico Fermi proposed a mechanism for the mysterious origin
of acceleration of the cosmic rays (Fermi 1949). He suggested that
it is the process of collisions with intra-galactic giant moving
magnetic fields that accelerates cosmic rays.

The accelerators based on the original idea of Enrico Fermi display
rich dynamical behavior both in the classical and the quantum
evolution. In 1961, Ulam studied the classical dynamics of a
particle bouncing on a surface which oscillates with a certain
periodicity. The dynamics of the bouncing particle is bounded by a
fixed surface placed parallel to the oscillating surface (Ulam
1961).

In Fermi-Ulam accelerator model, the presence of classical chaos
(Lieberman 1972, Lichtenberg 1980, 1983, 1992) and quantum chaos
(Karner 1989, Seba 1990) has been proved. A comprehensive work has
been devoted to study the classical and quantum characteristics of
the system (Lin 1988, Makowski 1991, Reichl 1992, Dembi\'nski 1995).
However, a particle bouncing in this system has a limitation, that,
it does not accelerate forever.

Thirty years after the first suggestion of Fermi, Pustyl'nikov
provided detailed study of another accelerator model. He replaced
the fixed horizontal surface of Fermi-Ulam model by a gravitational
field. Thus, Pustyl'nikov considered the dynamics of a particle on a
periodically oscillating surface in the presence of a gravitational
field (Pustyl'nikov 1977).

In his work, he proved that a particle bouncing in the accelerator
system finds modes, where it ever gets unbounded acceleration. This
feature makes the Fermi-Pustyl'nikov model richer in dynamical
beauties. The schematic diagram of Fermi-Ulam and Fermi-Pustyl'nikov
model is shown in Fig~\ref{fg:fam}.

In case of the Fermi-Ulam model, the absence of periodic
oscillations of reflecting surface makes it equivalent to a particle
bouncing between two fixed surfaces. However, in case of the
Fermi-Pustyl'nikov model, it makes the system equivalent to a ball
bouncing on a fixed surface under the influence of gravity. These
simple systems have thoroughly been investigated in classical and
quantum domains (Langhoff 1971, Gibbs 1975, Desko 1983, Goodins
1991, Whineray 1992, Seifert 1994, Bordo 1997, Andrews 1998,
Gea-Banacloche 1999).

In the presence of an external periodic oscillation of the
reflecting surface, the Fermi-Ulam model and the Fermi-Pustyl'nikov
model display the minimum requirement for a system to be chaotic
(Lichtenberg 1983, 1992). For the reason, these systems set the
stage to understand the basic characteristics of the classical and
quantum chaos (Jos\'e 1986, Seba 1990, Badrinarayanan 1995, Mehta
1990, Reichl 1986). Here, we focus our attention mainly on the
classical and quantum dynamics in the Fermi-Pustyl'nikov model. For
the reason, we name it as Fermi accelerator model in the rest of the
report.

\begin{figure}[t]
\begin{center}
\end{center}
\caption{a) Schematic diagram of the Fermi-Ulam accelerator model: A
particle moves towards a periodically oscillating horizontal
surface, experiences an elastic collision, and bounces off in the
vertical direction. Later, it bounces back due to another fixed
surface parallel to the previous one. b) Schematic diagram of the
Fermi-Pustyl'nikov model: A particle observes a bounded dynamics on
a periodically oscillating horizontal surface in the presence of a
constant gravitational field, $g$. The function, f(t), describes the
periodic oscillation of the horizontal surfaces.} \label{fg:fam}
\end{figure}

\subsection{Classical and quantum chaos in atom optics}

Quantum chaos, as the study of the quantum characteristics of the
classically chaotic systems, got immense attention after the work of
Bayfield and Koch on microwave ionization of hydrogen (Bayfield and
Koch, 1974). In the system the suppression of ionization due to the
microwave field was attributed to dynamical localization (Casati et
al.,1987;Koch and Van Leeuwen 1995). Later, the phenomenon was
observed experimentally (Galvez 1988, Bl\"umel 1989, Bayfield 1989,
Arndt 1991, Benvenuto 1997, Segev 1997).

Historically, the pioneering work of Giulio Casati and co-workers on
delta kicked rotor unearthed the remarkable property of dynamical
localization of quantum chaos (Casati 1979). They predicted that a
quantum particle exhibits diffusion following classical evolution
till quantum break time. Beyond this time the diffusion stops due to
quantum interference effects. In 1982, Fishman, Grempel and Prange
proved mathematically that the phenomenon of the dynamical
localization in kicked rotor is the same as Anderson localization of
solid state physics (Fishman 1982).

The study of quantum chaos in atom optics began with a proposal by
Graham, Schlautmann, and Zoller (Graham 1992). They investigated the
quantum characteristics of an atom which passes through a phase
modulated standing light wave. During its passage the atom
experiences a momentum transfer by the light field. The classical
evolution in the system exhibits chaos and the atom displays
diffusion. However, in the quantum domain, the momentum distribution
of the atom at the exit is exponentially localized or dynamically
localized, as shown in figure~\ref{zoller}.

\begin{figure}[t]
\begin{center}
\end{center}
\caption{Dynamical localization of an atom in a phase modulated
standing wave field: The time averaged probability distribution in
momentum space displays the exponentially localized nature of
momentum states. The vertical dashed lines give the border of the
classically chaotic domain. The exponential behavior on
semi-logarithmic plot is shown also by the dashed lines. (Graham
{\it et al.} 1992).} \label{zoller}
\end{figure}

Experimental study of quantum chaos in atom optics is largely based
upon the work of Mark Raizen (Raizen 1999). In a series of
experiments he investigated the theoretical predictions regarding
the atomic dynamics in a modulated standing wave field and regarding
delta kicked rotor model. The work also led to the invention of
newer methods of atomic cooling as well (Ammann 1997).

In the framework of atom optics, periodically driven systems have
been explored to study the characteristics of the classical and
quantum chaos. These systems include an atom in a modulated
electro-magnetic standing wave field (Graham 1992, Raizen 1999), an
ion in a Paul trap in the presence of an electromagnetic field
(Ghafar 1997), an atom under the influence of strong electromagnetic
pulses (Raizen 1999), and an atom in a Fermi accelerator (Saif 1998,
2000, 2000a, 2002).

The atom optics Fermi accelerator is advantageous in many ways: It
is analogous to the problem of a hydrogen atom in a microwave field.
In the absence of external modulation, both the systems possess
weakly binding potentials for which level spacing reduces with
increase in energy. However, Fermi accelerator is more promising due
to the absence of inherent continuum in the unmodulated case as it
is found in the hydrogen atom.

For a small modulation strength and in the presence of a low
frequency of the modulation, an atom exhibits bounded and integrable
motion in the classical and quantum domain (Saif 1998). However, for
larger values of the strength and/or the higher frequency of the
modulation, there occurs classical diffusion. In the corresponding
quantum domain, an atom displays no diffusion in the Fermi
accelerator, and eventually displays exponential localization {\it
both} in the position space and the momentum space. The situation
prevails till a critical value of the modulation strength which is
based purely on quantum laws.

The quantum delocalization (Chirikov 1986) of the matter waves in
the Fermi accelerator occurs at higher values of strength and
frequency of the modulation, above the critical value. The
transition from dynamical localization to quantum delocalization
takes place as the spectrum of the Floquet operator displays
transition from pure point to quasi-continuum spectrum (Brenner
1996, Oliveira 1994, Benvenuto 1991).

In nature, interference phenomena lead to revivals (Averbukh 1989a,
Averbukh 1989b, Alber 1990, Fleischhauer 1993, Chen 1995, Leichtle
1996a, Leichtle 1996b, Robinett 2000). The occurrence of the revival
phenomena in time dependent systems has been proved to be their
generic property (Saif 2005), and regarded as a test of
deterministic chaos in quantum domain (Bl\"umel 1997). The atomic
evolution in Fermi accelerator displays revival phenomena as a
function of the modulation strength and initial energy of the atom
(Saif 2000, 2000c, 2000d, 2000e, 2002).

\subsection{Layout}

This report is organized as follows: In section~\ref{int}, we review
the interaction of an atom with an optical field. In
section~\ref{Atomoptics}, we briefly summarize the essential ideas
of the quantum chaos dealt within the report. The experimental
progress in developing atom optics elements for the atomic de
Broglie waves is discussed in section~\ref{sec:atofacc}. These
elements are the crucial ingredients of complex atom optics systems,
as explained in section~\ref{complex}. We discuss the
spatio-temporal characteristics of the classical chaotic evolution
in section~\ref{cd}. The Fermi accelerator model is considered as
the focus of our study. In the corresponding quantum system the
phenomenon of dynamical localization is discussed in
section~\ref{qd}. The study of the recurrence phenomena in general
periodically driven systems is presented in section~\ref{dynrev},
and their study in the Fermi accelerator is made in
section~\ref{dynrevf}. In section~\ref{sum}, we make a discussion on
chaotic dynamics in periodically driven atom optics systems. We
conclude the report, in section~\ref{dedecoh}, by a brief discussion
of decoherence in quantum chaotic systems.


\section{Interaction of an atom with an optical field}
\label{int}

In this section, we provide a review of the steps leading to the
effective Hamiltonian which governs atom-field interaction. We are
interested in the interaction of a strongly detuned atom with a
classical light field. For the reason, we provide quantum mechanical
treatment to the center of mass motion however develop
semi-classical treatment for the interaction of the internal degrees
of freedom and optical field.

For a general discussion of the atom-field interaction, we refer to
the literature (Cohen-Tannoudji 1977, Kazantsev 1990, Meystre 2001,
Shore 1990, Sargent 1993, Scully 1997, Schleich 2001).

\subsection{Interacting atom}

We consider a two-level atom with the ground and the excited states
as $|g\rangle $ and $|e\rangle $, respectively. These states are
eigen states of the atomic Hamiltonian, $H_0$, and the corresponding
eigen energies are $\hbar \omega _{(g)}$ and $\hbar \omega _{(e)}$.
Thus, in the presence of completeness relation, $|g\rangle \langle
g|+|e\rangle \langle e|={\bf 1}$, we may define the atomic
Hamiltonian, as
\begin{equation}
H_{0}=\hbar \omega _{(e)}|e\rangle \langle e|+\hbar \omega
_{(g)}|g\rangle \langle g|.
\end{equation}

In the presence of an external electromagnetic field, a dipole is
formed between the electron, at position ${\bf r}_e$, and the
nucleus of the atom at, ${\bf r}_n$. In relative coordinates the
dipole moment becomes  $e{\bf r}_o=e({\bf r}_e-{\bf r}_n)$.
Consequently, the interaction of the atom with the electromagnetic
field is governed by the interaction Hamiltonian,
\begin{equation}
H_{int}=-e{\bf r}_o\cdot E({\bf r}+\delta{\bf r},t).
\end{equation}
Here, ${\bf r}$ is the center-of-mass position vector and
$\delta{\bf r}$ is either given by $\delta{\bf r}=-m_e{\bf
r}/(m_e+m_n)$ or $\delta{\bf r}=m_n{\bf r}/(m_e+m_n)$. The symbols
$m_e$ and $m_n$ define mass of the electron and that of nucleus,
respectively.

\subsubsection{Dipole Approximation}

Keeping in view a comparison between the wavelength of the
electromagnetic field and the atomic size, we consider that the
field does not change significantly over the dimension of the atom
and take $E({\bf r}+\delta{\bf r},t)= E({\bf r},t)$. Thus, the
interaction Hamiltonian becomes,
\begin{equation}
H_{int}\approx - e{\bf r}_o\cdot E({\bf r},t).
\end{equation}

In the analysis we treat the atom quantum mechanically, hence , we
express the dipole in operator description, as ${\hat{\vec{\wp}}}$.
With the help of the completeness relation, we may define the dipole
operator, as
\begin{eqnarray}
{\hat{\vec{\wp}}}= {\bf 1}\cdot{\hat{\vec{\wp}}}\cdot {\bf 1} =
e(|g\rangle\langle g| +|e\rangle\langle e|){\bf r}_o
(|g\rangle\langle g|+ |e\rangle\langle e|) = {\vec{\wp}}
|e\rangle\langle g|+ {\vec{\wp}}^*|g\rangle\langle e|.
\end{eqnarray}
Here, we have introduced the matrix element, $\vec{\wp}\equiv
e\langle e|{\bf r_o}|g\rangle$, of the electric dipole moment.
Moreover, we have used the fact that the diagonal elements, $\langle
g|{\bf r_o}|g\rangle$ and $\langle e|{\bf r_o}|e\rangle$, vanish as
the energy eigen states have well-defined parity.

We define $\hat{\sigma}^{\dag}\equiv |e\rangle\langle g|$ and
$\hat{\sigma}\equiv |g\rangle\langle e|$, as the atomic raising and
lowering operators, respectively. In addition, we may consider the
phases of the states $|e\rangle$ and $|g\rangle$ such that the
matrix element ${\vec{\wp}}$ is real (Kazantsev 1990). This leads us
to represent the dipole operator, as
\begin{equation}
{\hat{\vec{\wp}}}={\vec{\wp}}(\hat{\sigma} + \hat{\sigma}^{\dag}).
\end{equation}
Therefore, we arrive at the interaction Hamiltonian
\begin{equation}
{\hat H}_{int}=
-(\hat{\sigma}^{\dag} + \hat{\sigma}){\vec{\wp}}\cdot {\bf E}({\bf r},t).
\end{equation}

Thus, the Hamiltonian which descries the interaction of the atom
with the optical field in the presence of its center-of-mass motion,
reads as
\begin{equation}
{\hat H}=\frac{{\hat{\bf p}}^{2}}{2m}+\hbar \omega _{(e)}|e\rangle
\langle e|+\hbar \omega _{(g)}|g\rangle \langle g|-
(\hat{\sigma}^{\dag}+\hat{\sigma}) {\vec{\wp}}\cdot {\bf
E}({\hat{\bf r}},t), \label{Hamiltonian}
\end{equation}
where, ${\hat{\bf p}}$ describes the center-of-mass momentum. Here,
the first term on the right hand side corresponds to the kinetic
energy. It becomes crucial when there is a significant atomic motion
during the interaction of the atom with the optical field.

\subsection{Effective potential}

The evolution of the atom in the electromagnetic field becomes time
dependent in case the field changes in space and time, such that
\begin{equation}
{\bf E}({\bf r},t)=E_0\, {\bf u}({\bf r}) \cos\omega_f t.
\label{fieldt}
\end{equation}
Here, $E_0$ and $\omega_f$ are the amplitude and the frequency of
the field, respectively. The quantity ${\bf u}({\bf r})={\bf e}_0\,
u({\bf r})$ denotes the mode function, where ${\bf e}_0$ describes
the polarization vector of the field.

\subsubsection{Rotating wave approximation}

The time dependent Schr\"{o}dinger equation,
\begin{equation}
\emph{i}\hbar \frac{\partial |\psi \rangle }{\partial t}={\hat
H}|\psi \rangle, \label{eq:gtdsch}
\end{equation}
controls the evolution of the atom in the time dependent
electromagnetic field. Here, ${\hat H}$, denotes the total
Hamiltonian, given in equation~(\ref{Hamiltonian}), in the presence
of field defined in equation (\ref{fieldt}).

In order to solve the time dependent Schr\"{o}dinger equation, we
express the wave function, $|\psi \rangle$, as an ansatz, i.e.
\begin{equation}
|\psi \rangle \equiv e^{-\emph{i}\omega _{(g)}t}\psi _{g}({\bf
r},t)|g\rangle +e^{-\emph{i}(\omega_{(g)}+\omega_f )t}\psi _{e}({\bf
r},t)|e\rangle . \label{ansatz}
\end{equation}
Here, $\psi _{g}=\psi _{g}({\bf r},t)$ and $\psi _{e}=\psi _{e}({\bf
r},t)$, express the probability amplitudes in the ground state and
excited state, respectively.

We substitute equation~(\ref{ansatz}), in the time dependent
Schr\"{o}dinger equation, given in equation~(\ref{eq:gtdsch}). After
a little operator algebra we, finally get the coupled equations for
the probability amplitudes, $\psi _{g}$ and $\psi _{e}$, expressed
as
\begin{eqnarray}
\emph{i}\hbar \frac{\partial \psi _{g}}{\partial t} =\frac{\hat{\bf
p}^2}{2m}
\psi _{g}-\frac{1}{2}\hbar \Omega_R%
u({\hat{\bf r}})(1+e^{-2\emph{i}\omega_f t})\psi _{e}, \label{rate1}\\
\emph{i}\hbar \frac{\partial \psi _{e}}{\partial t} =\frac{\hat{\bf
p}^2}{2m} \psi _{e} -\hbar \delta \psi _{e}-\frac{1}{2}\hbar
\Omega_R u({\hat{\bf r}})(1+e^{-2\emph{i}\omega_f t})\psi _{g}.
\label{rate2}
\end{eqnarray}
Here, $\Omega_R \equiv {\vec{\wp}}\cdot {\bf e}_0 E_{0}/2\hbar $,
denotes the Rabi frequency, and $\delta \equiv \omega_f -\omega
_{eg}$ is the measure of the tuning of the external field frequency,
$\omega_f$, away from the atomic transition frequency, expressed as
$\omega _{eg}= \omega _{(e)}-\omega _{(g)}$.

In the rotating wave approximation, we eliminate the rapidly
oscillating terms in the coupled equations~(\ref{rate1}) and
(\ref{rate2}) by {\it averaging} them over a period, $\tau \equiv
2\pi/\omega_f$. We assume that the probability amplitudes, $\psi
_{e}$ and $\psi _{g}$, do not change appreciably over the time scale
and approximate,
\begin{equation}
1+e^{-2\emph{i}\omega_f t}\sim 1.
\end{equation}
Thus, we eliminate the explicit time dependence in the equations
(\ref{rate1}) and (\ref{rate2}), and get
\begin{eqnarray}
\emph{i}\hbar \frac{\partial \psi _{g}}{\partial t} = \frac{{\hat
{\bf p}}^2}{2m}\psi_{g} -\hbar \Omega_R u({\hat {\bf r}})\psi_{e},
\label{eq:eqprm1}\\
\emph{i}\hbar \frac{\partial \psi _{e}}{\partial t} = \frac{{\hat
{\bf p}}^2}{2m} \psi_{e}-\hbar \delta \psi _{e}-\hbar\Omega_R
u({\hat {\bf r}})\psi _{g}.  \label{eq:eqprm2}
\end{eqnarray}

\subsubsection{Adiabatic approximation}

In the limit of a large detuning between the field frequency,
$\omega_f$, and the atomic transition frequency $\omega_{eg}$,  we
consider that the excited state population changes adiabatically.
Hence, we take $\psi_{e}(t)\approx \psi_{e}(0)=constant$.

As a consequence, in equation (\ref{eq:eqprm2}) the deviation of the
probability amplitude $\psi _{e}$ with respect to time and position
becomes vanishingly small. This provides an approximate value for
the probability amplitude $\psi_e(t)$, as
\begin{equation}
\psi_e(t)\approx -\frac{\Omega_R}{\delta} u({\hat {\bf r}})\psi
_{g}(t).
\end{equation}

Thus, the time dependent Schr\"odinger equation for the ground state
probability amplitude, $\psi_g$, becomes
\begin{equation}
i\hbar \frac{\partial \psi }{\partial t}\cong \frac{{\hat {\bf
p}}^2}{2 m} \psi+\frac{\hbar\Omega^2_R}{\delta}u^2({\hat{\bf
r}})\psi . \label{eff}
\end{equation}
For simplicity, here and later in the report, we drop subscript $g$
and take $\psi_{g}\equiv\psi $.

The Schr\"odinger equation, given in equation~(\ref{eff}),
effectively governs the dynamics of a super-cold atom, in the
presence of a time dependent optical field. The atom almost stays in
its ground state under the condition of a large atom-field detuning.

\subsubsection{The Effective Hamiltonian}
\label{sec:melse}

Equation~(\ref{eff}) leads us to an effective Hamiltonian,
$H_{eff}$, such as
\begin{equation}
H_{eff}\equiv \frac{{\hat {\bf p}}^2}{2 m}
+\frac{\hbar\Omega^2_R}{\delta}u^2({\hat{\bf r}}). \label{haeff}
\end{equation}
Here, the first term describes the center-of-mass kinetic energy,
whereas, the second term indicates an effective potential as seen by
the atom. The spatial variation in the potential enters through the
mode function of the electromagnetic field, $u$. Fascinatingly, we
note that by a proper choice of the mode function almost any
potential for the atom can be made.

The second term in equation (\ref{haeff}), leads to an effective
force $F=-2(\hbar\Omega^2/\delta)u\bigtriangledown u$, where
$\bigtriangledown u$ describes gradient of $u$. Hence, the
interaction of the atom with the electromagnetic field exerts a
position dependent gradient force on the atom, which is directly
proportional to the {\it square} of the Rabi frequency,
$\Omega^2_R$, and inversely proportional to the detuning, $\delta$.


\subsection{Scattering atom}

In the preceding subsections we have shown that an atom in a
spatially varying electromagnetic field experiences a position
dependent effective potential. As a consequence the atom with a
dipole moment gets deflected as it moves through an optical field
(Moskowitz 1983) and experiences a position dependent gradient
force. The force is the largest where the gradient is the largest.
For example, an atom in a standing light field experiences a maximum
force at the nodes and at antinodes it observes a vanishing dipole
force as the gradient vanishes.

We enter into a new regime, when the kinetic energy of the atom
parallel to the optical field is comparable to the recoil energy. We
consider the propagation of the atom in a potential formed by the
interaction between the dipole and the field, moreover, we treat the
center-of-mass motion along the cavity quantum mechanically.


\section{Quantum characteristics of chaos}
\label{Atomoptics}

Over last twenty five years the field of atom optics has matured and
become a rapidly growing branch of quantum electronics. The study of
atomic dynamics in a phase modulated standing wave by Graham,
Schlautman and Zoller (Graham 1992), made atom optics a testing
ground for quantum chaos as well. Their work got experimental
verification as the dynamical localization of cold atoms was
observed later  in the system in momentum space (Moore 1994,
Bardroff 1995, Amman 1997).

\subsection{Dynamical localization}

In 1958, P. W. Anderson showed the absence of diffusion in certain
random lattices (Anderson 1958). His distinguished work was
recognized by a Nobel prize, in 1977, and gave birth to the
phenomenon of localization in solid state physics --- appropriately
named as Anderson localization (Anderson 1959).

An electron, on a one dimensional crystal lattice displays
localization if the equally spaced lattice sites are taken as
random. The randomness may arise due to the presence of impurities
in the crystal. Thus, at each $i$th site of the lattice, there acts
a random potential, $T_i$. The probability amplitude of the hopping
electron, $u_i$, is therefore expressed by the Schr\"odinger
equation
\begin{equation}
T_i u_i + \sum_{r\neq 0} W_r u_{i+r} = E u_i, \label{ander}
\end{equation}
where, $W_r$ is the hopping amplitude from $i$th site to its $r$th
neighbor. If the potential, $T_i$, is periodic along the lattice,
the solution to the equation~(\ref{ander}) is Bloch function, and
the energy eigen values form a sequence of continuum bands. However,
in case $T_i$ are uncorrelated from site to site, and distributed
following a distribution function, eigen values of the
equation~(\ref{ander}) are exponentially localized. Thus, in this
situation the hopping electron finds itself localized.

Twenty years later, Giulio Casati and coworkers suggested the
presence of a similar phenomenon occurring in the kicked rotor model
(Casati 1979). In their seminal work they predicted that a quantum
particle, subject to periodic kicks of varying strengths, exhibits
the suppression of classical diffusion. The phenomenon was named as
dynamical localization. Later, on mathematical grounds, Fishman,
Grempel and Prange developed equivalence between dynamical
localization and the Anderson localization (Fishman 1982).

In its classical evolution an ensemble of particles displays
diffusion as it evolves in the kicked rotor in time. In contrast,
the corresponding quantum system follows classical evolution only up
to a certain time. Later, it displays dynamical localization as the
suppression of the diffusion or its strong reduction (Brivio 1988,
Izrailev 1990).

The maximum time for which the quantum dynamics mimics the
corresponding classical dynamics is the quantum break time of the
dynamical system (Haake 1992, Bl\"umel 1997). Before the quantum
break time, the quantum system follows classical evolution and fills
the stable regions of the phase space (Reichl 1986).

The phenomenon of dynamical localization is a fragile effect of
coherence and interference. The quantum suppression of classical
diffusion does not occur under the condition of resonance or in the
presence of some translational invariance (Lima 1991, Guarneri
1993). The suppression of diffusion in the absence of these
conditions is dominantly because of restrictions of quantum dynamics
by quantum cantori (Casati 1999a, Casati 1999b).

The occurrence of dynamical localization is attributed to the change
in statistical properties of the spectrum (Berry 1981, Zaslavsky
1981, Haake 1992, Altland 1996). The basic idea involved is that,
integrability corresponds to Poisson statistics (Brody 1981, Bohigas
1984), however, non-integrability corresponds to Gaussian orthogonal
ensemble (GOE) statistics as a consequence of the Wigner's level
repulsion.

In the Fermi-Ulam accelerator model, for example, the quasi energy
spectrum of the Floquet operator displays such a transition. It
changes from Poisson statistics to GOE statistics as the effective
Planck's constant changes in value (Jos{\'e} 1986).

The study of the spectrum leads to another interesting understanding
of localization phenomenon. Under the influence of external periodic
force, as the system exhibits dynamical localization the spectrum
changes to a pure point spectrum (Prange 1991, Jos\'{e} 1991, Dana
1995). There occurs a phase transition to a quasi continuum spectrum
leading to quantum delocalization (Benvenuto 1991, Oliveira 1994),
for example, in the Fermi accelerator.

\subsection{Dynamical recurrences}

In one dimensional bounded quantum systems the phenomena of quantum
recurrences exist (Robinett 2000). The roots of the phenomena are in
the quantization or discreteness of the energy spectrum. Therefore,
their occurrence provides a profound manifestation of quantum
interference.

In the presence of an external time dependent periodic modulation,
these systems still exhibit the quantum recurrence phenomena (Saif
2002, Saif 2005). Interestingly, their occurrence is regarded as a
manifestation of deterministic quantum chaos (Bl\"umel 1997).

Hogg and Huberman provided the first numerical study of quantum
recurrences in chaotic systems by analyzing the kicked rotor model
(Hogg 1982). Later, the recurrence phenomena was further
investigated in various classically chaotic systems, such as, the
kicked top model (Haake 1992), in the dynamics of a trapped ion
interacting with a sequence of standing wave pulses (Breslin 1997),
in the stadium billiards (Tomsovic 1997) and in multi-atomic
molecules (Grebenshchikov 1997).

The phenomena of quantum recurrences are established generic to
periodically driven quantum systems which may exhibit chaos in their
classical domain (Saif 2002, Saif 2005). The times of quantum
recurrences are calculated by secular perturbation theory, which
helps to understand them as a function of the modulation strength,
initial energy of the atom, and other parameters of the system (Saif
2000b, Saif 2000c).

In a periodically driven system, the occurrence of recurrences
depends upon the initial conditions in the phase space. This helps
to analyze quantum nonlinear resonances and the quantum stochastic
regions by studying the recurrence structures (Saif 2000c).
Moreover, it is suggested that the quantum recurrences serve as a
very useful probe to analyze the spectrum of the dynamical systems
(Saif 2005).

In the periodically driven one-dimensional systems, the spectra of
Floquet operators serve as quasi-energy spectra of the time
dependent systems. A lot of mathematical work has been devoted to
the study of Floquet spectra of periodically driven systems (Breuer
1989, 1991a, 1991b).

It has been established that, in the presence of external modulation
in the tightly binding potentials where the level spacing between
adjacent levels increases with the increase in energy, the quasi
energy spectrum remains a point spectrum regardless of the strength
of the external modulation (Hawland 1979, 1987, 1989a, 1989b, Joye
1994). However, in the presence of external modulation in weakly
binding potentials, where the level spacing decreases with increase
in energy it changes from a point spectrum to a continuum spectrum
above a certain critical modulation strength (Delyon 1985, Benvenuto
1991, Oliveira 1994, Brenner and Fishman 1996). In tightly binding
potentials, the survival of point spectrum reveals quantum
recurrences, whereas, the disappearance of point spectrum in weakly
binding potentials leads to the disappearance of quantum recurrences
above a critical modulation strength (Saif 2000c, Iqbal 2005).

\subsection{Poincare' recurrences}

According to the Poincare' theorem a trajectory always return to a
region around its origin, but the statistical distribution depends
on the dynamics. For a strongly chaotic motion, the probability to
return or the probability to survive decays exponentially with time
(Lichtenberg 1992). However, in a system with integrable and chaotic
components the survival probability decays algebraically (Chirikov
1999). This change in decay rate is subject to the slowing down of
the diffusion due to chaos border determined by some invariant tori
(Karney 1983, Chirikov 1984, Meiss 1985, Meiss 1986, Chirikov 1988,
Ruffo 1996).

The quantum effects modify the decay rate of the survival
probability (Tanabe 2002). It is reported that in a system with
phase space a mixture of integrable and chaotic components, the
algebraic decay $P(t)\sim 1/t^p$ has the exponent $p=1$. This
behavior is suggested to be due to tunneling and localization
effects (Casati 1999c).


\section{Mirrors and cavities for atomic de Broglie waves}
\label{sec:atofacc}

As discussed in section~\ref{int}, by properly tailoring the spatial
distribution of an electromagnetic field we can create almost any
potential we desire for the atoms. Moreover, we can make the
potential repulsive or attractive by making a suitable choice of the
atom-field detuning. An atom, therefore, experiences a repulsive
force as it interacts with a blue detuned optical field, for which
the field frequency is larger than the transition frequency.
However, there is an attractive force on the atom if it finds a red
detuned optical field, which has a frequency smaller than the atomic
transition frequency. Thus, in principle, we can construct any atom
optics component and apparatus for the matter waves, analogous to
the classical optics.

\subsection{Atomic mirror}

A mirror for the atomic de Broglie waves is a crucial ingredient of
the atomic cavities. An atomic mirror is obtained by an
exponentially decaying optical field or an evanescent wave field.
Such an optical field exerts an exponentially increasing repulsive
force on an approaching atom, detuned to the blue (Bordo 1997).

How to generate an evanescent wave field is indeed an interesting
question. In order to answer this question, we consider an
electromagnetic field $E({\bf r},t)= E({\bf r})e^{-i\omega_f t}$,
which travels in a dielectric medium with a dielectric constant, $n$
and undergoes total internal reflection. The electromagnetic field
inside the dielectric medium reads
\begin{equation}
E({\bf r},t)= E_0 e^{i{\bf{k}\cdot\bf{r}}-i\omega_f t}{\bf e_r}.
\end{equation}
where, $\bf e_r$ is the polarization vector and ${\bf k}=k{\hat k}$
is the propagation vector.

The electromagnetic field, $E({\bf r},t)$, is incident on an
interface between the dielectric medium with the dielectric
constant, $n$, and another dielectric medium with a smaller
dielectric constant, $n_1$. The angle of incidence of the field is
$\theta_i$ with the normal to the interface.

Since the index of refraction $n_1$ is smaller than $n$, the angle
$\theta_r$ at which the field refracts in the second medium, is
larger than $\theta_i$. As we increase the angle of incidence
$\theta_i$, we may reach a critical angle, $\theta_i=\theta_c$, for
which $\theta_r=\pi/2$. According to Snell's law, we define this
critical angle of incidence as
\begin{equation}
\theta_c\equiv\sin^{-1}\left(\frac{n_1}{n}\right).
\end{equation}

Hence, for an electromagnetic wave with an angle of incidence larger
than the critical angle, that is, $\theta_{i} > \theta_{c}$, we find
the inequality, $\sin \theta _{r}> 1$ (Mandel 1986, Mandel and Wolf
1995). As a result, we deduce that $\theta _{r}$ is imaginary, and
define
\begin{equation}
\cos \theta_{r}=i\sqrt{\left( \frac{\sin \theta_{i}}{\sin \theta
_{c}}\right) ^{2}-1}.
\end{equation}
Therefore, the field in the medium of smaller refractive index,
$n_{1}$, reads
\begin{eqnarray}
E({\bf r},t) ={\bf e_r}E_{0}e^{ik_{1}x\sin \theta _{r}+ik_{1}z\cos
\theta _{r}}e^{-i\omega_f t} ={\bf e_r}E_{0}e^{-\kappa z}e^{i(\beta
x-\omega_f t)} \label{ert}
\end{eqnarray}
where, $\kappa =k_{1}\sqrt{(\sin \theta _{i}/\sin \theta
_{c})^{2}-1}$ and $\beta =k_{1}\sin \theta _{i}/\sin \theta _{c}$
(Jackson 1965). Here, $k_{1}$ defines the wave number in the medium
with the refractive index $n_{1}$. This demonstrates that in case of
total internal reflection the field along the normal of the
interface decays in the positive $z$ direction, in the medium with
the smaller refractive index.

In 1987, V. Balykin and his coworkers achieved the first
experimental realization of an atomic mirror (Balykin 1987). They
used an atomic beam of sodium atoms incident on a parallel face
plate of fused quartz and observed the specular reflection.

They showed that at small glancing angles, the atomic mirror has a
reflection coefficient equal to unity. As the incident angle
increases a larger number of atoms reaches the surface and undergoes
diffusion. As a result the reflection coefficient decreases. The
reflection of atoms bouncing perpendicular to the mirror is
investigated in reference (Aminoff 1993) and from a rough atomic
mirror studied in reference (Henkel 1997).


\subsubsection{Magnetic mirror}

We can also construct atomic mirror by using magnetic fields instead
of optical fields. At first, magnetic mirror was realized to study
the reflection of neutrons (Vladimirski\^{i} 1961). In atom optics,
the use of a magnetic mirror was suggested in reference (Opat 1992),
and later it was used to study the reflection of incident rubidium
atoms perpendicular to the reflecting surface (Roach 1995, Hughes
1997a, Hughes 1997b, Saba 1999). Recently, it has become possible to
modulate the magnetic mirror by adding a time dependent external
field. We may also make controllable corrugations which can be
varied in a time shorter than the time taken by atoms to interact
with the mirror (Rosenbusch 2000a, Rosenbusch 2000b).

Possible mirror for atoms is achieved by means of surface plasmons
as well (Esslinger 1993, Feron 1993, Christ 1994). Surface plasmons
are electromagnetic charge density waves propagating along a
metallic surface. Traveling light waves can excite surface plasmons.
The technique provides a tremendous enhancement in the evanescent
wave decay length (Esslinger 1993).


\subsection{Atomic cavities}

Based on the atomic mirror various kinds of atomic cavities have
been suggested. A system of two atomic mirrors placed at a distance
with their exponentially decaying fields in front of each other form
a cavity or resonator for the de Broglie waves. The atomic cavity is
regarded as an analog of the Fabry Perot cavity for radiation fields
(Svelto 1998).

By using more than two mirrors, other possible cavities can be
developed as well. For example, we may create a ring cavity for the
matter waves by combining three atomic mirrors (Balykin 1989).

An atomic gravitational cavity is a special arrangement. Here, atoms
move under gravity towards an atomic mirror, made up of an
evanescent wave (Matsudo 1997). The mirror is placed perpendicular
to the gravitational field. Therefore, the atoms observe a normal
incidence with the mirror and bounce back. Later, they exhaust their
kinetic energy moving against the gravitational field (Kasevich
1990, Wallis 1992) and return. As a consequence, the atoms undergo a
bounded motion in this atomic trampoline or atomic gravitational
cavity. Hence, the evanescent wave mirror together with the
gravitational field constitutes a cavity for atoms.


\subsection{Gravitational cavity}

The dynamics of atoms in the atomic trampoline or atomic
gravitational cavity attracted immense attention after the early
experiments by the group of S. Chu at Stanford, California (Kasevich
1990). They used a cloud of sodium atoms stored in a magneto-optic
trap and cooled down to 25$\mu$K.

As the trap switched off the atoms approached the mirror under
gravity and display a normal incidence. In their experiments two
bounces of the atoms were reported. The major noise sources were
fluctuations in the laser intensities and the number of initially
stored atoms. Later experiments reported up to thousand bounces
(Ovchinnikov 1997).

Another kind of gravitational cavity can be realized by replacing
the optical evanescent wave field by liquid helium, forming the
atomic mirror for hydrogen atoms. In this setup hydrogen atoms are
cooled below $0.5K^{o}$ and a specular reflection of $80\%$ has been
observed (Berkhout 1989).

\subsubsection{A bouncing atom}

An atom dropped from a certain initial height above an atomic mirror
experiences a linear gravitational potential,
\begin{equation}
V_{gr}=mgz,
\end{equation}
as it approaches the mirror. Here, $m$ denotes mass of the atom, $g$
expresses the constant gravitational acceleration and $z$ describes
the atomic position above the mirror. Therefore, by taking the
optical field given in Eq.~\ref{ert} and using Dipole approximation
and rotating wave approximation the effective Hamiltonian, given in
equation~(\ref{haeff}), becomes
\begin{equation}
H_{eff}=\frac{p^2}{2m}+mgz+\hbar\Omega_{eff}e^{-2\kappa z}.
\label{eq:cohaml}
\end{equation}
The effective Hamiltonian governs the center-of-mass motion of the
atom in the presence of the gravity above evanescent wave field.
Here, $\Omega_{eff}=\Omega^2_R/\delta$, describes the effective Rabi
frequency, and $\kappa^{-1}$ describes the decay length of the
atomic mirror.

The optical potential is dominant for smaller values of $z$ and
decays exponentially as $z$ becomes larger. Thus, for larger
positive values of $z$ as the optical potential approaches zero the
gravitational potential takes over, as shown in figure~\ref{fg:gcav}
(left).

The study of the spatio-temporal dynamics of the atom in the
gravitational cavity explains interesting dynamical features. In the
long time dynamics the atom undergoes self interference and exhibits
revivals, and fractional revivals. Furthermore, as a function of
space and time, the quantum interference manifests itself
interestingly in the quantum carpets (Gro{\ss}mann 1997, Marzoli
1998), as shown in figure~\ref{fg:gcav} (right).

We show the quantum carpets for an atom in the Fermi accelerator, in
figure~\ref{fg:gcav}. The dark gray regions express the larger
probabilities, whereas, the in between light gray regions indicate
smaller probabilities to find the atom in its space-time evolution.
We recognize these regular structures not related to the classical
space-time trajectories. Close to the surface of the atomic mirror,
at $z=0$, these structures appear as vertical canals sandwiched
between two high probability dark gray regions. These canals become
curved gradually away from the atomic mirror where the gravitational
field is significant.

\begin{figure}[t]
\begin{center}
\end{center}
\caption{(left) As we switch off the
Magneto-optic trap (MOT) at time, ${t}=0$, the atomic wave packet
starts its motion from an initial height. It moves under the
influence of the linear gravitational potential $V_{\rm gr}$ (dashed
line) towards the evanescent wave atomic mirror. Close to the
surface of the mirror the effect of the evanescent light field is
dominant and the atom experiences an exponential repulsive optical
potential $V_{\rm opt}$ (dashed line). Both the potentials together
make the gravitational cavity for the atom (solid line). (right) We
display the space-time quantum mechanical probability distribution
of the atomic wave packet in the Fermi accelerator represented as a
quantum carpet.} \label{fg:gcav}
\end{figure}

\subsubsection{Mode structure}

An atom observes almost an instantaneous impact with the atomic
mirror, as it obeys two conditions: {\it (i)} The atom is initially
placed away from the atomic mirror in the gravitational field, and
{\it (ii)} the atomic mirror is made up of an exponentially decaying
optical field with a very short decay length. Thus, we may take the
gravitational cavity in a triangular well potential, made up of a
linear gravitational potential and an infinite potential ,where the
atom undergoes a bounded motion

We may express the corresponding effective Hamiltonian as
\begin{equation}
H=\frac{p^{2}}{2m}+V(z),
\end{equation}
where
\begin{equation}
V(z)\equiv \left\{
\begin{array}{r@{\quad \quad}l}
mgz \,\,\,\,\,\, z\ge 0\,,\\
\infty \,\,\,\,\,\, z< 0\,.
\end{array}
\right.  \label{eq:vapp}
\end{equation}
The solution to the stationary Schr\"{o}dinger equation, $H\psi_n
=E_n\psi_n$, hence provides
\begin{equation}
\psi_{n}={\cal N}\,Ai(z-z_{n}) \label{airy}
\end{equation}
as the eigen function (Wallis 1992). Here, ${\cal N}$ expresses the
normalization constant. Furthermore, we take $x\equiv(2m^{2}g/\hbar
^{2})^{1/3}z$ as the dimensionless position variables, and
$z_n=(2m^{2}g/\hbar ^{2})^{1/3}z_{n}$ as the $n$th zero of the Airy
function. The index $n$ therefore defines the $n$th mode of the
cavity. Here the, $n$th energy eigen value is expressed as
\begin{equation}
E_{n}=(\frac{1}{2}m\hbar ^{2}g^{2})^{1/3}z_{n}=mgz_{n}.
\end{equation}

In figure \ref{modest}, we show the Wigner distribution,
\begin{equation}
W(z, p)=\frac{1}{2\pi \hbar}\int_{-\infty}^{\infty}\psi_n^*(z+y/2)
\,\psi_n(z-y/2) \,e^{i\frac{p}{2\hbar}y}\,dy,
\end{equation}
for the first three eigen-functions, {\it i.e.} $n=1, 2, 3$, in the
scaled coordinates, $z$, and $p$. The distributions are symmetric
around $p=0$ axes. Moreover, we note that the distributions extends
in space along, $z$-axis as $n$ increases. We can easily identify
the non-positive regions of the Wigner distribution functions, as
well.

\begin{figure}[t]
\begin{center}
\end{center}
\caption{We express the Wigner distribution function for the first
three eigen functions, defined in Eq.~(\ref{airy}).}
\label{modest}
\end{figure}


\subsection{Optical traps}

A slight change of the shape of the atomic mirror from flat to
concave helps to make successful trap for atoms in a gravitational
cavity (Wallis 1992, Ovchinnikov 1995). In addition, atomic
confinement is possible by designing bi-dimensional light traps on a
dielectric surface (Desbiolles 1996) or around an optical fiber (Fam
Le Kien 2005).

By appropriate choice of attractive and/or repulsive evanescent
waves, we can successfully trap or guide atoms in a particular
system. In the presence of a blue detuned optical field and another
red detuned optical field with a smaller decay length, a net
potential is formed on or around a dielectric surface. The
bi-dimensional trap can be used to store atoms, as shown in figure
\ref{trap}.

We may make an optical cylinder within a hollow optical fiber to
trap and guide cold atoms (Renn 1995, Ito 1996). A laser light
propagating in the glass makes an evanescent wave within the fiber.
The optical field is detuned to the blue of atomic resonance, thus
it exerts repulsive force on the atoms, leading to their trapping at
the center of the fiber. The system also serves as a useful
waveguide for the atoms.

\begin{figure}[t]
\begin{center}
\end{center}
\caption{(left) Schematic diagram of a bi-dimensional atomic trap
around an optical fiber. (right) Transverse plane profile of the
total potential, $U_{tot}$, produced by net optical potential and
the van der Waals potential. (Fam Le Kien {\it et al.} 2004).}
\label{trap}
\end{figure}


\section{Complex systems in atom optics}
\label{complex}

In atom optics systems, discussed in section \ref{sec:atofacc}, the
atomic dynamics takes place in two dimensions. However in the
presence of an external driving force on the atoms an explicit time
dependence appears in these systems. The situation may arise due to
a phase modulation and/or amplitude modulation of the optical field.
We may make the general Hamiltonian description of these driven
systems as,
\begin{equation}
H=H_0(x,p)- V(t)u(x+\varphi(t)). \label{hcom}
\end{equation}
Here, $H_0$ controls atomic dynamics in the absence of driving force
and along $x$-axis. Furthermore, $V(t)$ and $\varphi(t)$ are
periodic functions of time, and $u$ mentions the functional
dependence of the potential in space.

Equation (\ref{hcom}) reveals that the presence of explicit time
dependence in these systems provides a three dimensional phase
space. Hence, in the presence of coupling, these systems fulfill the
minimum criteria to expect chaos. Various situations have been
investigated and important understandings have been made regarding
the atomic evolution in such complex systems. Following we make a
review of the atom optics systems studied in this regard.

\subsection{Phase modulated standing wave field}

In 1992, Graham, Schlautmann and Zoller (GSZ) studied the dynamics
of an atom in a monochromatic electromagnetic standing wave field
made up of two identical and aligned counter propagating waves. As
one of the running waves passes through an electro-optic modulator,
a phase modulation is introduced in the field (Graham 1992).

The effective Hamiltonian which controls the dynamics of the atom
is,
\begin{equation}
H=\frac{p^2}{2m}- V_0\cos(kx+\varphi(t)),
\end{equation}
where, $V_0$ expresses the constant amplitude and
$\varphi=x_0\sin(\Omega t)$ defines the phase modulation of the
field, with a frequency $\Omega$ and amplitude $x_0$.

According to GSZ model, the atom in the phase modulated standing
wave field experiences random kicks. Thus, a momentum is transferred
from the modulated field to the atom along the direction of the
field.

In classical domain the atom exhibits classical chaos as a function
of the strength of the phase modulation and undergoes diffusive
dynamics. However, in the corresponding quantum dynamics the
diffusion is sharply suppressed and the atom displays an
exponentially localized distribution in the momentum space (Moore
1994, Bardroff 1995).

\subsection{Amplitude modulated standing wave field}

The atomic dynamics alters as the atom moves in an amplitude
modulated standing wave field instead of a phase modulated standing
wave field. The amplitude modulation may be introduced by providing
an intensity modulation to the electromagnetic standing wave field
through an acousto-optic modulator.

The effective Hamiltonian which controls the atomic dynamics can now
be expressed as
\begin{equation}
H=\frac{p^2}{2m}- V_0\cos(\Omega t)\cos(kx).
\end{equation}
Thus, the system displays a double resonance structure as
\begin{equation}
H=\frac{p^2}{2m}- \frac{V_0}{2}[\cos(kx+\Omega t)+ \cos(kx-\Omega
t)].
\end{equation}
The atom, hence, finds two primary resonances at $+\Omega$ and
$-\Omega$, where it rotates clockwise or counter clockwise with the
field (Averbuckh 1995, Gorin 1997, Monteoliva 1998).

\subsection{Kicked rotor model}
\label{krm}

The group of Mark Raizen at Austin, Texas presented the experimental
realization of the Delta Kicked Rotor in atom optic (Robinson 1995,
Raizen 1999). This simple system is considered as a paradigm of
chaos (Haake 2001). In their experiment a cloud of ultra cold sodium
atoms experiences a one dimensional standing wave field which is
switched on instantaneously, and periodically after a certain period
of time.

The standing light field makes a periodic potential for the atoms.
The field frequency is tuned away from the atomic transition
frequency. Therefore, we may ignore change in the probability
amplitude of any excited state as a function of space and time in
adiabatic approximation. The general Hamiltonian which effectively
governs the atomic dynamics in the ground state therefore becomes,
\begin{equation}
H=\frac{p^2}{2m}- V_0\cos kx.
\end{equation}
Here, the amplitude $V_0$ is directly proportional to the intensity
of the electromagnetic field and inversely to the detuning.

The simple one dimensional system may become non-integrable and
display chaos as the amplitude of the spatially periodic potential
varies in time. The temporal variations are introduced as a train of
pulses, each of a certain finite width and appearing after a
definite time interval, $T$. Thus, the complete Hamiltonian of the
driven system appears as,
\begin{equation}
H=\frac{p^2}{2m}- V_0\cos kx\sum_{n=-\infty}^{+\infty}\delta(t-nT).
\end{equation}
The atom, therefore, experiences a potential which displays spatial
as well as temporal periodicity.

It is interesting to note that between two consecutive pulses the
atom undergoes free evolution, and at the onset of a pulse it gets a
kick which randomly changes its momentum. The particular system,
thus, provides an atom optics realization of Delta Kicked Rotor and
enables us to study the theoretical predictions in a quantitative
manner in laboratory experiments  .

\subsection{Triangular billiard}


When a laser field makes a triangle in the gravitational field the
atoms may find themselves in a triangular billiard. An interesting
aspect of chaos enters depending on the angle between the two sides
forming the billiard. Indeed the atomic dynamics is ergodic if the
angle between the laser fields is an irrational multiple of $\pi$
(Artuso 1997a, Artuso 1997b) and may be pseudo integrable if the
angles are rational (Richens 1981).

\subsection{An Atom optics Fermi accelerator}
\label{atfa}

The work horse of the Fermi accelerator is the gravitational cavity.
In the atomic Fermi accelerator, an atom moves under the influence
of gravitational field towards an atomic mirror made up of an
evanescent wave field. The atomic mirror is provided a spatial
modulation by means of an acousto-optic modulator which provides
intensity modulation to the incident laser light field (Saif 1998).

Hence, the ultra cold two-level atom, after a normal incidence with
the modulated atomic mirror, bounces off and travels in the
gravitational field, as shown in figure~\ref{fg:model}. In order to
avoid any atomic momentum along the plane of the mirror the laser
light which undergoes total internal reflection, is reflected back.
Therefore, we find a standing wave in the plane of the mirror which
avoids any specular reflection (Wallis 1995).

The periodic modulation in the intensity of the evanescent wave
optical field may lead to the spatial modulation of the atomic
mirror as
\begin{equation}
I(z,t) = I_0\exp[-2\kappa z +\epsilon\sin(\Omega t)].
\label{eq:modin}
\end{equation}
Thus, the center-of-mass motion of the atom in z- direction follows
effectively from the Hamiltonian
\begin{eqnarray}
\tilde{H}=\frac{{p}_{z}^{2}}{2m}+m g z+\hbar \Omega_{eff}
\exp[-2\kappa z+\epsilon\sin(\Omega t)], \label{eq:ham}
\end{eqnarray}
where, $\Omega _{{eff}}$ denotes the effective Rabi frequency.
Moreover, $\epsilon$ and $\Omega$ express the amplitude and the
frequency of the external modulation, respectively. In the absence
of the modulation the effective Hamiltonian, given in
equation~(\ref{eq:ham}), reduces to equation~(\ref{eq:cohaml}).

In order to simplify the calculations, we may make the variables
dimensionless by introducing the scaling, ${z}\equiv (\Omega^2/g)z$,
${p}\equiv (\Omega/mg){p}_{z}$ and ${t}\equiv \Omega t$. Thus, the
Hamiltonian becomes,
\begin{equation}
H=\frac{{ p}^2}{2}+{ z} +V_0\exp[-\kappa ({ z}-\lambda\sin { t})]
\;, \label{eq:scham}
\end{equation}
where, we take $H=(\Omega^2/mg^2)\tilde{H}$,
$V_0=\hbar\Omega_{eff}\Omega^2/mg^2$, $\kappa={\kappa} g/\Omega^2$,
and $\lambda=\epsilon\Omega^2/g$.

\begin{figure}[t]
\begin{center}
\end{center}
\caption{A cloud of atoms is trapped and cooled in a
magneto-optic trap (MOT) to a few micro-Kelvin.
The MOT is placed at a certain height above a dielectric slab.
An evanescent wave created by the total internal reflection of the
incident laser beam from the surface of the dielectric
serves as a mirror for the atoms. At the onset of the experiment
the MOT is switched off and the atoms move under gravity towards the
exponentially decaying evanescent wave field. Gravity and the evanescent
wave field form a cavity for the atomic de Broglie waves. The atoms undergo
bounded motion in this gravitational cavity.
An acousto-optic modulator provides
spatial modulation of the evanescent wave field.
This setup serves as a realization of an atom optics Fermi accelerator.}
\label{fg:model}
\end{figure}


\section{Classical chaos in Fermi accelerator}
\label{cd}

The understanding of the classical dynamics of the Fermi accelerator
is developed together with the subject of classical chaos (Lieberman
1972, Lichtenberg 1980, 1983, 1992). The dynamics changes from
integrable to chaotic and to accelerated regimes in the accelerator
as a function of the strength of modulation. Thus, a particle in the
Fermi accelerator exhibits a rich dynamical behavior.

In the following, we present a study of the basic characteristics of
the Fermi accelerator.


\subsection{Time evolution}
\label{claps}

The classical dynamics of a single particle bouncing in the Fermi
accelerator is governed by the Hamilton's equations of motion. The
Hamiltonian of the system, expressed in Eq.~(\ref{eq:scham}), leads
to the equations of motion as
\begin{eqnarray}
\dot{{ z}}=\frac{\partial H}{\partial { p}} = { p}\,,  \label{eq:ham1} \\
\dot{{ p}}=-\frac{\partial H}{\partial { z}} = -1 + \kappa V_0
e^{-\kappa({ z}-\lambda \sin { t})}\,.  \label{eq:ham2}
\end{eqnarray}

In the absence of external modulation, the equations (\ref{eq:ham1})
and (\ref{eq:ham2}) are nonlinear and the motion is regular, with no
chaos (Langhoff 1971, Gibbs 1975, Desko 1983, Goodins 1991, Whineray
1992, Seifert 1994, Bordo 1997, Andrews 1998, Gea-Banacloche 1999).
However, the presence of an external modulation introduces explicit
time dependence, and makes the system suitable for the study of
chaos.

In order to investigate the classical evolution of a statistical
ensemble in the system, we solve the Liouville equation (Gutzwiller
1992). The ensemble comprises a set of particles, each defined by an
initial condition in phase space. Interestingly, each initial
condition describes a possible state of the system.

We may represent the ensemble by a distribution function,
$P_0({z_0,p_0})$ at time $t=0$ in position and momentum space. The
distribution varies with the change in time, $t$. Hence, at any
later time $t$ the distribution function becomes, $P(z,p,t)$.

In a conservative system the classical dynamics obeys the condition
of incompressibility of the flow (Lichtenberg 1983), and leads to
the Liouville equation, expressed as,
\begin{equation}
\left\{\frac{\partial}{\partial { t}} +
{ p}\frac{\partial}{\partial { z}} +
\dot{{ p}}
\frac{\partial}{\partial { p}} \right\} P({z,p,t})
= 0\;.  \label{eq:liouv}
\end{equation}
Here, $\dot{{ p}}$ is the force experienced by every particle of the
ensemble in the Fermi accelerator, defined in
equation~(\ref{eq:ham2}). Hence, the classical Liouville equation
for an ensemble of particles becomes,
\begin{equation}
\left\{ \frac{\partial}{\partial { t}} +
{ p} \frac{\partial}{\partial { z}}
-[1-\kappa V_0 \exp\{-\kappa({ z}-\lambda \sin{ t})\}]
\frac{\partial}{\partial { p}}\right\}
P({z,p,t}) = 0\;.  \label{eq:reliou}
\end{equation}

The general solution of equation~(\ref{eq:reliou}) satisfies the
initial condition, $P(z,p,t=0)=P_0({ z}_0,{ p}_0)$. According to the
method of characteristics (Kamke 1979), it is expressed as
\begin{equation}
P({ z,p,t})=\int_{-\infty}^{\infty} dz_0 \int_{-\infty}^{\infty}
dp_0 \delta\{{ z}-\bar{{z}}({ z}_0,{ p}_0,{ t})\} \delta\{{
p}-\bar{{ p}}({ z}_0,{ p}_0,{ t})\} P_0({ z}_0,{ p}_0)\;.
\label{kamk}
\end{equation}
Here, the classical trajectories $\bar{{ z}}=\bar{{ z}}({ z}_0,{
p}_0,{ t})$ and $\bar{{ p}}= \bar{{ p}}({ z}_0,{ p}_0,{ t})$ are the
solutions of the Hamilton equations of motion, given in
equations~(\ref{eq:ham1}) and (\ref{eq:ham2}). This amounts to say
that each particle from the initial ensemble follows the classical
trajectory $(\bar{{ z}}, \bar{{ p}})$. As the system is
nonintegrable in the presence of external modulation, we solve
equation~(\ref{kamk}) numerically.


\subsection{Standard mapping}
\label{sm}

We may express the classical dynamics of a particle in the Fermi
accelerator by means of a mapping. The mapping connects the momentum
of the bouncing particle and its phase just before a bounce to the
momentum and phase just before the previous bounce. This way the
continuous dynamics of a particle in the Fermi accelerator is
expressed as discrete time dynamics.

In order to write the mapping, we consider that the modulating
surface undergoes periodic oscillations following sinusoidal law.
Hence, the position of the surface at any time is $z=\lambda\sin t$,
where $\lambda$ defines the modulation amplitude. In the scaled
units the time of impact, $t$, is equivalent to the phase $\varphi$.

Furthermore, we consider that the energy and the momentum remain
conserved before and after a bounce and the impact is elastic. As a
result, the bouncing particle gains twice the momentum of the
modulated surface, that is, $2\lambda\cos\varphi$, at the impact.
Here, we consider the momentum of the bouncing particle much smaller
than that of the oscillating surface. Moreover, it undergoes an
instantaneous bounce.

Keeping these considerations in view, we express the momentum,
$p_{{\rm i}+1}$, and the phase, $\varphi_{{\rm i}+1}$, just before
the $({\rm }i+1)$th bounce in terms of momentum, $p_{\rm i}$, and
the phase, $\varphi_{\rm i}$, just before the ${\rm i}$th bounce, as
\begin{eqnarray}
p_{{\rm i}+1}=-p_{\rm i} -\Delta t_{\rm i} +2\lambda\cos\varphi_{\rm i}\;,  \nonumber \\
\varphi_{{\rm i}+1}= \varphi_{\rm i} + \Delta t_{\rm i}\;.
\label{eq:tdmap}
\end{eqnarray}
Here, the time interval, $\Delta t_{\rm i}\equiv t_{{\rm
i}+1}-t_{\rm i}$, defines the time of flight between two consecutive
bounces. Hence, the knowledge of the momentum and the phase at the
${\rm i}$th bounce leads us to the time interval $\Delta t_{\rm i}$
as the roots of the equation,
\begin{equation}
p_{\rm i} \Delta t_{\rm i} - \frac{1}{2}\Delta t_{\rm i}^2=
\lambda(\sin(\varphi_{\rm i}+\Delta t_{\rm i})-\sin\varphi_{\rm
i})\,.
\end{equation}

We consider that the amplitude of the bouncing particle is large
enough compared to the amplitude of the external modulation,
therefore, we find no kick-to-kick correlation. Thus, we may assume
that the momentum of the particle just before a bounce is equal to
its momentum just after the previous bounce. The assumptions permit
us to take, $\Delta t_{\rm i}\sim -2p_{{\rm i}+1}$. As a result the
mapping reads,
\begin{eqnarray}
p_{{\rm i}+1}= p_{\rm i} - 2\lambda\cos\varphi_{\rm i}\;,  \nonumber \\
\varphi_{{\rm i}+1}= \varphi_{\rm i} - 2 p_{{\rm i}+1}\;.
\label{eq:chirmap}
\end{eqnarray}

We redefine the momentum as ${\wp}_i=-2p_{\rm i}$, and $K=4\lambda$.
The substitutions translate the mapping to the standard
Chirikov-Taylor mapping (Chirikov 1979), that is,
\begin{eqnarray}
{\wp}_{{\rm i}+1}= {\wp}_{\rm i} +K\cos\varphi_{\rm i}\;,  \nonumber \\
\varphi_{{\rm i}+1}= \varphi_{\rm i} + {\wp}_{{\rm i}+1}\;.
\label{eq:chimap1}
\end{eqnarray}
Hence, we can consider the Fermi accelerator as a discrete dynamical
system (Kapitaniak 2000).

The advantage of the mapping is that it depends only on the kick
strength or chaos parameter, $K=4\lambda$. Therefore, simply by
changing the value of the parameter, $K$, the dynamical system
changes from, stable with bounded motion to chaotic with unbounded
or diffusive dynamics. There occurs a critical value of the chaos
parameter, at which the change in the dynamical characteristics
takes place, is $K=K_{cr}=0.9716....$ (Chirikov 1979, Greene 1979).

\subsection{Resonance overlap}
\label{res}

Periodically driven systems, expressed by the general Hamiltonian
given in equation~(\ref{hcom}), exhibit resonances. These resonances
appear whenever the frequency of the external modulation, $\Omega$,
matches with the natural frequency of unmodulated system, $\omega$
(Lichtenberg 1992, Reichl 1992). Thus, the resonance condition
becomes,
\begin{equation}
N\omega-M\Omega=0,  \label{eq:reson}
\end{equation}
where, $N$ and $M$ are relative prime numbers. These resonances are
spread over the phase space of the dynamical systems.

Chirikov proved numerically that in a discrete dynamical system
expressed by Chirikov mapping, the resonances remain isolated so far
as the chaos parameter, $K$, is less than a critical value,
$K_{cr}=1$ (Chirikov 1979). Later, by his numerical analysis, Greene
established a more accurate measure for the critical chaos parameter
as $K_{cr} \simeq 0.9716...$ (Greene 1979). Hence, the dynamics of a
particle in the system remains bounded in the phase space by
Kolmogorov-Arnold-Moser (KAM) surfaces (Arnol'd 1988, Arnol'd 1968).
As a result, only local diffusion takes place.

Following our discussion presented in the section~\ref{sm}, we note
that in the Fremi accelerator the critical chaos parameter,
$\lambda_l$, is defined as $\lambda_l\equiv K_{cr}/4\simeq 0.24$.
Hence, for a modulated amplitude much smaller than $\lambda_l$ the
phase space is dominated by the invariant tori, defining KAM
surfaces. These surfaces separate resonances. However, as the
strength of modulation increases, area of the resonances grow
thereby more and more KAM surfaces break.

At the critical modulation strength, $\lambda_l$, last KAM surface
corresponding to golden mean is broken (Lichtenberg 1983). Hence,
above the critical modulation strength the driven system has no
invariant tori and the dynamics of the bouncing particle is no more
restricted, which leads to a global diffusion.

The critical modulation strength $\lambda_{l}$, therefore, defines
an approximate boundary for the onset of the classical chaos in the
Fermi accelerator. The dynamical system exhibits bounded motion for
a modulation strength $\lambda$ smaller than the critical value
$\lambda_l$, and a global diffusion beyond the critical value.


\subsection{Brownian motion}
\label{brm}

As discussed in section \ref{sm}, classical dynamics of a particle
in the Fermi accelerator is expressed by the standard mapping. This
allows to write the momentum as
\begin{equation}
{\wp}_j= {\wp}_0+4\lambda \sum_{n=1}^j\cos\varphi_n, \label{eq:momi}
\end{equation}
at the $j$th bounce. Here, ${\wp}_0$ is the initial scaled momentum.
In order to calculate the dispersion in the momentum space with the
change of modulation strength, $\lambda$, we average over the phase,
$\varphi$. This yields,
\begin{equation}
\Delta {\wp_j}^2 \equiv \langle {\wp_j}^2\rangle-\langle
{\wp_j}\rangle^2= (4\lambda)^2\sum_{n=1}^j\sum_{n^{\prime}=1}^j
\langle\cos(\varphi_n)\cos(\varphi_{n^{\prime}})\rangle\,.
\label{eq:depis}
\end{equation}

As the amplitude of the bouncing particle becomes very large as
compared with that of the oscillating surface, we consider that no
kick-to-kick correlation takes place. This consideration leads to a
random phase for the bouncing particle in the interval $[0,2\pi]$ at
each bounce above the critical modulation strength, $\lambda_l$.
Thus a uniform distribution of the phase appears over many bounces.
Therefore, we get,
\begin{equation}
\langle\cos(\varphi_n)\cos(\varphi_{n^{\prime}})\rangle=\langle\cos^2\varphi_n\rangle
\delta_{n,n^{\prime}}=\frac{1}{2}\delta_{n,n^{\prime}}\;.
\label{eq:cosf}
\end{equation}
Hence, equation (\ref{eq:depis}) provides
\begin{equation}
\Delta p^2=2j\,\lambda^2=j\,D\;,  \label{eq:linp}
\end{equation}
where $D=2\lambda^2$ and expresses the diffusion constant. Moreover,
$j$ describes the number of bounces.

Since we can find an average period over $j$ bounces, the number of
bounces grows linearly with time. This leads to a linear growth of
the square of width in momentum space as a function of the evolution
time, as we find in the Brownian motion.

It is conjectured that in long time limit the diffusive dynamics of
the Fermi accelerator attains Boltzmann distribution. We may express
the distribution as,
\begin{equation}
P_{cl}(z,p)=(2\pi)^{-1/2}\eta^{-3/2}\exp[ -(p^2/2+ z)/\eta]\;,
\label{eq:dist}
\end{equation}
where, the quantity $\eta$ represents effective temperature (Saif
1998).

Hence, the momentum distribution follows Maxwellian distribution in
the classical phase space, and is Gaussian, that is,
\begin{equation}
P_{cl}(p)=\frac{1}{\sqrt{2\pi\eta}}\exp[-p^2/(2\eta)]\;.
\label{eq:distp}
\end{equation}
Moreover, the classical position distribution follows exponential
barometric formula
\begin{equation}
P_{cl}(z)=\frac{1}{\eta}\exp(-z/\eta)\;,  \label{eq:distz}
\end{equation}
predicted from Eq.~(\ref{eq:dist}). The conjecture is confirmed from
the numerical study, as we shall see in section~\ref{probdist}. The
equations~(\ref{eq:distp}) and (\ref{eq:distz}) lead to the
conclusion that $\Delta p^2 =\Delta z= jD$.

\subsection{Area preservation}
\label{area}

Preservation of area in phase space is an important property of the
Hamiltonian systems (Pustyl'nikov 1977, Lichtenberg 1983) and
corresponds to the conservation of energy. In a system the
determinant of Jacobian, defined as,
\begin{equation}
{\cal J}\equiv det\left(
\begin{array}{cc}
\frac{\partial\varphi_{{\rm i}+1}}{\partial\varphi_{\rm i}} \,\,\,
\frac{\partial\varphi_{{\rm i}+1}%
}{\partial {\wp}_{\rm i}} \\
\frac{\partial {\wp}_{{\rm i}+1}}{\partial\varphi_{\rm i} } \,\,\,
\frac{\partial
{\wp}_{{\rm i}+1}}{%
\partial {\wp}_{\rm i}}
\end{array}
\right)\;,  \label{eq:jacob}
\end{equation}
leads to the verification of area preservation in a classical
system.

In the case of a non-dissipative, conservative system the
determinant of the Jacobian is equal to one which proves the
area-preserving nature of the system in phase space. However, for a
dissipative system it is less than one.

For the Fermi accelerator the determinant of the Jacobian, obtained
with the help of equation~(\ref{eq:chimap1}), is
\begin{equation}
{\cal J}=det\left(
\begin{array}{cc}
1-K\sin\varphi_{{\rm i}} \,\,\, 1 \\
-K\sin\varphi_{{\rm i}} \,\,\, 1
\end{array}
\right) =1\;.  \label{eq:jac}
\end{equation}
Hence, the dynamics of a particle in the Fermi accelerator exhibits
the property of area preservation in phase space.


\subsection{Lyapunov exponent}
\label{lyap}

The exponential sensitivity of a dynamical system on the initial
conditions is successfully determined by means of Lyapunov exponent.
The measure of Lyapunov exponent is considered to determine the
extent of classical chaos in a chaotic system (Schuster 1989,
Gutzwiller 1992, Ott 1993).

We may define the exponent as,
\begin{equation}
{\cal L}= \lim_{t\rightarrow\infty}\,\frac{1}{t}\,
log\left(\frac{d(t)}{d(0)} \right).  \label{eq:lexp}
\end{equation}
Here, $d(0)$ represents distance between two phase points at time,
$t=0$, and $d(t)$ describes their distance after an evolution time,
$t$. If the dynamics is regular and non diffusive, the exponent will
be zero. However if it is diffusive we find nonzero positive
Lyapunov exponent. A system displaying dissipative dynamics has a
nonzero negative exponent.

For weak modulation, that is for $\lambda < \lambda_l$, we find the
exponent zero in general. This indicates a dominantly regular
dynamics in the Fermi accelerator for weak modulation strength. For
modulation strengths $\lambda>\lambda_l$, the dynamics is dominantly
diffusive in the system established by nonzero positive exponent
(Saif 1998).


\subsection{Accelerating modes}
\label{am}

The classical work of Pustyl'nikov on the Fermi accelerator model
(Pustyl'nikov 1977) guarantees the existence of a set of initial
data, such that, trajectories which originate from the set always
speed-up to infinity. The initial data appears always in circles of
radius $\rho> 0$ and has positive Lebesgue measure. Furthermore, the
existence of these accelerating modes requires the presence of
certain windows of modulation strength, $\lambda$.

The windows of modulation strength which supports accelerated
trajectories, read as,
\begin{equation}
s\pi \leq \lambda < \sqrt{1+(s\pi)^2}\;.  \label{eq:lcon}
\end{equation}
Here, $s$ takes integer and half integer values for the sinusoidal
modulation of the reflecting surface.

In order to probe the windows of modulation strength which support
the unbounded acceleration, we may calculate the width of momentum
distribution, $\Delta p\equiv \sqrt{\langle{p^2}\rangle-{\langle
p\rangle}^2}$, numerically as a function of the modulation strength.
Here, $\langle p\rangle$ and $\langle{p^2}\rangle$ are the first and
the second moment of momentum, respectively.

We consider an ensemble of particles which is initially distributed
following Gaussian distribution.
In order to study the dynamical behavior of the ensemble we record
its width in the momentum space after a propagation time for
different modulation strengths. We note that for very small
modulation strengths, the widths remain small and almost constant,
which indicates no diffusive dynamics, as shown in
figure~\ref{fg:pres}. For larger values of $\lambda$ the width
$\Delta p$ increases linearly, which follows from the equation
(\ref{eq:linp}).

We find that at the modulation strengths which correspond to the
windows on modulation strengths, given in equation (\ref{eq:lcon}),
the diffusion of the ensemble is at its maximum. The behavior occurs
as the trajectories which correspond to the areas of phase space
supporting accelerated dynamics undergo coherent acceleration,
whereas the rest of the trajectories of the initial distribution
display maximum diffusion. For the reason, as the modulation
strengths increase beyond these values, the dispersion reduces as we
find in Fig.~\ref{fg:pres}.

Equation~(\ref{eq:lcon}) leads to the modulation strength,
$\lambda_m$, for which maximum accelerated trajectories occur in the
system. We find these values of $\lambda_m$ as,
\begin{equation}
\lambda_{m}= \frac{s\pi + \sqrt{1+(s\pi)^2}}{2}\;.  \label{eq:lm}
\end{equation}
The value is confirmed by the numerical results (Saif 1999), as
expressed in figure~\ref{fg:pres}.

For a modulation strength given in equation~(\ref{eq:lcon}) and an
initial ensemble originating from the area of phase space which
supports accelerated dynamics, we find sharply suppressed value of
the dispersion. It is a consequence of a coherent acceleration of
the entire distribution (Yannocopoulos 1993, Saif 1999).

\begin{figure}
\begin{center}
\end{center}
\caption{The width of the momentum distribution $\Delta p$ is
plotted as a function of the
         modulation strength, $\lambda$.
         An ensemble of particles, initially in a Gaussian distribution,
         is propagated for a time, $t=500$. The numerical results depict the presence of
         accelerated dynamics for the modulation strengths, expressed in equation~(\ref{eq:lcon}).}
\label{fg:pres}
\end{figure}


\section{Quantum dynamics of the Fermi accelerator}
\label{qd}

The discrete time dynamics of a classical particle in the Fermi
accelerator is successfully described by the standard mapping. For
the reason, we find onset of chaos in the system as the chaos
parameter exceeds its critical value. The presence of classical
chaos in the Fermi accelerator is established by non-zero positive
Lyapunov exponent (Saif 1998). This makes the dynamics of a quantum
particle in the Fermi accelerator quite fascinating in the frame
work of quantum chaos.

Similar to classical dynamics, we find various dynamical regimes in
the quantum evolution of the system as a function of modulation
strength. The quantum system however has another controlling
parameter, that is, the scaled Planck's constant. The commutation
relation between the dimensionless coordinates, $z$ and $p$, defined
for equation (\ref{eq:scham}), appear as
\begin{equation}
[z,p]=ik^{\hspace{-2.1mm}-}.
\end{equation}
This leads to the definition of the scaled Planck's constant as
$k^{\hspace{-2.1mm}-}=(\Omega/\Omega_0)^3$. The frequency $\Omega_0$
takes the value as $\Omega_0=(mg^2/\hbar)^{1/3}$. Thus, it is easily
possible to move from the semi-classical to pure quantum mechanical
regime simply by changing the frequency of the external modulation
alone.

The quantum dynamics of the center-of-mass motion of an atom in the
Fermi accelerator, follows from the time dependent Schr\"{o}dinger
equation,
\begin{equation}
ik^{\hspace{-2.1mm}-}\frac{\partial \psi }{\partial t}=\left[
\frac{p^{2}}{2}%
+z+V_{0}\exp \left[ -\kappa (z-\lambda \sin t)\right] \right] \psi
\;. \label{eq:dham}
\end{equation}
The equation (\ref{eq:dham}) describes the dynamics of a cold atom
which moves towards the modulated atomic mirror under the influence
of gravitational field and bounces off. In its long time evolution
the atom experiences large number of bounces.


A quantum particle, due to its nonlocality, always experiences the
modulation of the atomic mirror during its evolution in the Fermi
accelerator. However, the classical counterpart ''feels'' the
modulation of the mirror only when it bounces off. This contrast in
quantum and classical evolution leads to profound variations in the
dynamical properties in the two cases. In the next sections we study
the quantum characteristics of the Fermi accelerator.

\subsection{Near integrable dynamics} \label{dl}

As we discussed in section \ref{res}, for a modulation strength
$\lambda$ smaller than $\lambda_l$, classical resonances remain
isolated. The classical dynamics of a particle in the Fermi
accelerator is, therefore, restricted by KAM surfaces. As a result,
after its initial spread over the area of resonances, a classical
ensemble stops diffusing and the classical dynamics stays bounded.

Similar to the isolated classical resonances, isolated quantum
resonances prone to exist for smaller modulations (Berman and
Kolovsky 1983, Reichel and Lin 1986). For the reason an evolution
similar to the classical evolution is found even in the quantum
domain. We find this behavior as a wave packet in its quantum
dynamics mimics classical bounded motion, both in the position and
momentum space in the Fermi accelerator, for smaller modulations.

Manifestation of the behavior comes from the saturation of the
width, in the position space $\Delta z$ and momentum space $\Delta
p$, as a function of time, both in the classical and quantum domain.
The saturation values for the widths in the position space and the
momentum space may differ due to the size of underlying resonances
(Saif et al. 1998, Saif 1999).


\subsection{Localization window} \label{sec:locwin}

In the Fermi accelerator the classical system undergoes a global
diffusion as the resonance overlap takes place above the critical
modulation strength, $\lambda_l=0.24$. In contrast, in the
corresponding quantum mechanical domain there occurs another
critical modulation strength, $\lambda_u$, which depends purely on
the quantum laws.

At the critical modulation strength, $\lambda_u$, a phase transition
occurs and the quasi-energy spectrum of the Floquet operator changes
from a point spectrum to a continuum spectrum (Benvenuto 1991,
Oliveira 1994, Brenner 1996). We may define the critical modulation
strength for the Fermi accelerator as,
$\lambda_u\equiv\sqrt{k^{\hspace{-2.1mm}-}}/2$, when the exponential
potential of the atomic mirror is considered as an infinite
potential (Saif 1998). Beyond this critical modulation strength
quantum diffusion takes place.

Hence, the two conditions together make a window on the modulation
strength. We may find a drastic difference between the classical
dynamics and the corresponding quantum dynamics, for a modulation
strength which fulfills the condition,
\begin{equation}
\lambda_l<\lambda<\lambda_u\;.  \label{win}
\end{equation}

Within the window classical diffusion sets in whereas the
corresponding quantum dynamics displays localization. For the
reason, we name it as localization window (Chen 1997, Saif 1998,
Saif 1999).


\subsection{Beyond localization window}

Above the upper boundary of the localization window, $\lambda_u$,
even the quantum distributions display diffusion. For the reason, a
quantum wave packet maintains its widths, in the momentum and the
position space, only up to $\lambda \simeq \lambda_u$. Beyond the
value it starts spreading similar to the classical case.

For the reason, above the critical modulation strength, $\lambda_u$,
width in the momentum space $\Delta p$ and in the position space
$\Delta z$, display a growing behavior. However, in addition to an
overall growing behavior, there exist maxima and minima in a regular
fashion similar to the classical case, as the modulation strength
$\lambda$ increases. We show the classical behavior in
figure~\ref{fg:pres}.

The presence of maxima corresponds to maximum dispersion.
Interestingly, the maxima occur at $\lambda=\lambda_m$ as expressed
in equation~(\ref{eq:lm}), and the size of the peaks is determined
from equation~(\ref{eq:lcon}). Hence, we infer that the behavior is
a consequence of the accelerated trajectories, and in accordance
with the Pustyl'nikov's work, as discussed in section~\ref{am}.

For a modulation strength given in equation~(\ref{eq:lcon}), the
portions of a probability distribution which originate from the area
of phase space supporting accelerated dynamics, always get coherent
acceleration. However, the rest of the distribution displays maximum
diffusion and thus attains maximum widths.

The coherent acceleration manifests itself in the regular spikes in
the momentum and the position distributions, as shown in
figure~\ref{fg:dxdpl}. It is important to note that the spiky
behavior takes place for the modulation strengths which satisfy the
condition given in equation~(\ref{eq:lcon}).

These spikes gradually disappear as we choose the modulation
strength $\lambda$ away from these windows. The behavior is a
beautiful manifestation of the quantum non-dispersive accelerated
dynamics in the Fermi accelerator.

\begin{figure}[t]
\begin{center}
\end{center}
\caption{The classical and the quantum dynamics are compared far
         above the localization window
         by calculating the momentum distributions, $P(p)$, after the evolution time, $t=500$.
         (left) For $\lambda=1.7$
         and (right) for $\lambda=2.4$, marked in Fig.~\ref{fg:pres},
         we plot the momentum distributions in both the classical
         and the quantum space, together, as mirror images. We find spikes appearing in
         the momentum distributions for $\lambda=1.7$, which is due to the
         presence of coherent accelerated dynamics.}
\label{fg:dxdpl}
\end{figure}


\section{Dynamical localization of atoms}

The beautiful work of Casati and co-workers on the delta kicked
rotor, now a paradigm of classical and quantum chaos, led to the
discovery of dynamical localization (Casati 1979). They predicted
that a quantum particle follows diffusive dynamics of a classical
chaotic system only up to a certain time, named as quantum break
time. Beyond the time the classical diffusion is ceased due to the
quantum interference and the initial quantum distribution settles
into an exponential localization.

Similar behavior is predicted to occur in the quantum dynamics of an
atom in the Fermi accelerator (Chen 1997, Saif 1998). However, the
existence of the dynamical localization is restricted to the
localization window (Benvenuto 1991), defined in the
equation~(\ref{win}). Furthermore, the effective Planck's constant
imposes another condition as it is to be larger than a minimum
value, $4\lambda_l^2$.


\subsection{Probability distributions: An analysis}
\label{probdist}

An atomic wave packet, expressed initially as the ground state of
harmonic oscillator and satisfying minimum uncertainty criteria,
displays an exponential localization behavior when propagated in the
Fermi accelerator. The localization takes place both in the momentum
space and position space. Furthermore, it occurs for a time beyond
quantum break time, and for a modulation within the localization
window. The final quantum distributions are a manifestation of
interference phenomena, thereby independent of the choice of initial
distributions of the wave packet.

In the momentum space the atomic wave packet redistributes itself
around the initial mean momentum after an evolution time of many
bounces (Lin 1988). In the long time limit, the probability
distribution decays exponentially in the momentum space, as shown in
figure~\ref{fg:dpx}. It is estimated that the probability
distribution in momentum space follows, $\exp{(-|p|/\ell)}$,
behavior. Here, $\ell$ describes the localization length.

In the corresponding position space the wave packet reshapes itself
in a different manner. The mean position of the probability
distribution is estimated to be still at the initial mean position
of the wave packet. However, it decays exponentially in the
gravitational field when the exponent follows the square root law
(Saif 2000a), as shown in the figure~\ref{fg:dpx}.

Thence, the quantum distributions in the position space and in the
momentum space are completely different from their classical
counterparts, discussed in section 6.4. In contrast to the quantum
mechanical exponential localization in momentum space, classical
ensemble undergoes diffusion and attains Maxwell distribution,
defined in equation (\ref{eq:distp}).

Moreover in the position space the quantum distribution is
exponential with the exponent following square root behavior,
whereas, it follows exponential barometric equation in the classical
domain and the exponent follows linear behavior, as given in
equation~(\ref{eq:distz}). Thus, the quantum distribution displays
much rapid decay in the position space.

\begin{figure}
\begin{center}
\end{center}
\caption{Comparison between the classical (thick lines) and the
quantum mechanical (thin lines) distributions in momentum space
(left) and in position space (right) at a {\it fixed} evolution
time: We have $k^{\hspace{-2.1mm}-}=4$ for which the localization
window reads $0.24<\lambda<1$. We take the modulation strength,
$\lambda=0.8$, which is within the window. The initial minimum
uncertainty wave packet attains exponential distribution after a
scaled evolution time $t=3200$ in the Fermi accelerator, with the
exponents following linear and square root behavior in momentum
space and in position space, respectively. The corresponding
classical distributions follow Maxwell distribution and barometric
equation in momentum space and in position space, respectively.
Here, $\Delta z=2$ and the number of particles in the classical
simulation is 5000. The time of evolution, $t=3200$, is far above
the quantum break time $t^*=250$. The estimated behavior is
represented by thin dashed lines.} \label{fg:dpx}
\end{figure}


\subsection{Dispersion: Classical and quantum}

In the quantum dynamics of the delta kicked rotor model, energy
grows following the classical diffusive growth only up to quantum
break time, estimated as $t^*\approx
\lambda^2/{k^{\hspace{-2.1mm}-}}^2$. Later, it saturates and
deviates from the unlimited growth in energy which takes place in
the classical case.

In a dynamical system the suppression of classical diffusion in
quantum domain is regarded as an evidence of dynamical localization
(Reichl 1986). It is conjectured as independent of initial
distribution (Izrailev 1990).

G. J. Milburn discussed that a wave packet displays saturation in
its energy after quantum break time in the Fermi accelerator,
provided the modulation strength satisfies the condition given in
equation~(\ref{win}) (Chen and Milburn 1997). This saturation in
energy, when the corresponding classical system supports unlimited
gain, establishes the quantum suppression of the classical chaos in
the Fermi accelerator, as shown in figure \ref{fg:wpxm}.

\begin{figure}
\begin{center}
\end{center}
\caption{Comparison between classical (dashed line) and quantum
(solid line) dynamics {\it within} the localization window: The
classical and quantum evolution go together till the quantum break
time, $t^*$. Above the quantum break time, while the classical
ensemble diffuses as a function of time, the quantum system shows a
suppression of the diffusion. (Chen 1997).} \label{fg:wpxm}
\end{figure}

In the Fermi accelerator the phenomenon of dynamical localization
exists in the position and momentum space, as well. In the classical
dynamics of an ensemble in the Fermi accelerator, the dispersion in
the position space and the {\it square} of the dispersion in the
momentum space are calculated numerically. It is observed that they
exhibit a linear growth as a function of time.

It is straightforward to establish the behavior analytically as
$\Delta p^2 \sim \Delta z \sim Dt$, with the help of equations
(\ref{eq:distp}) and (\ref{eq:distz}). The corresponding quantum
mechanical quantities, however, saturate or display significantly
slow increase (Saif 1998, Saif 2000b).

\subsection{Effect of classical phase space on dynamical
localization}

As long as isolated resonances exist in the classical phase space,
the localization phenomenon remains absent in the corresponding
quantum dynamics. The dynamical localization however appears as the
overlap of the resonance takes place in the classical dynamics. As a
consequence in the quantum domain, the probability distributions
decay exponentially in the stochastic sea (Chirikov and Shepelyansky
1986).

In the Fermi accelerator the quantum probability distributions
display exponential localization. However, the distribution in
position space exhibits a different decay, as the exponent follows
square root behavior in contrast to a linear behavior as in the
momentum space (Saif et al. 1998, Saif 1999).

We note that in addition to an overall decay there occur plateau
structures in the two distributions both in the classical and in the
quantum dynamics. The numerical results provide an understanding to
the behavior as the confinement of the initial probability
distributions.

A portion of the initial probability distribution which originates
initially from a resonance finds itself trapped there. For the
reason over the area of a resonance the approximate distribution of
the probability is uniform which leads to a plateau (Chirikov 1986,
Saif 2000b).

In the presence of the numerical study we conjecture that in the
quantum domain corresponding to a classical resonance there occur a
discrete quasi-energy spectrum, whereas in the stochastic region we
find a quasi continuum of states (Saif 2000c). For the reason, the
quantum probability distribution which occupies the quasi-continuum
spectrum undergoes destructive interference and manifests the
exponential decay of the probability distributions.

The height of plateaus appearing in the probability distributions
rises for the higher values of the Planck's constant. However, their
size and the location remain the same (Lin 1988). Interestingly, for
the larger values of the effective Planck's constant,
$k^{\hspace{-2.1mm}-}$, the number of plateaus increases as well.

These characteristics are comprehensible as we note that, for a
larger values of $k^{\hspace{-2.1mm}-}$, the size of thee initial
minimum uncertainty wave packet increases. For the reason, the
portion of the initial probability distribution which enters the
stable islands increases, and therefore, displays an increase in the
height of the plateaus. Moreover, for a larger values of
$k^{\hspace{-2.1mm}-}$, the exponential tails of the quantum
distributions cover more resonances leading to more plateau
structures in the localization arms (Saif 2000b).


\subsection{Quantum delocalization}

Above the upper boundary of the localization window, the
interference is overall destructive and leads to the quantum
mechanical delocalization in the Fermi accelerator. Thus, we find
exponential localization within the localization window only, as
shown in figure~\ref{fg:mplocal1}. We may predict the presence of
quantum delocalization in a time dependent system by observing the
time independent spectrum of the potential.

In nature there exist two extreme cases of potentials: Tightly
binding potentials, for which the level spacing always increases
between adjacent energy levels as we go up in the energy, for
example, in one dimensional box; Weakly binding potentials, for
which the level spacing  always reduces between the adjacent levels,
and we find energy levels always at a distance smaller than before,
as we go up in the energy, for example, in the gravitational
potential. Quadratic potential offers a special case, as the level
spacing remains equal for all values of energy. Therefore, it lies
between the two kinds of potentials.

The fundamental difference of the time independent potentials has a
drastic influence on the properties of the system in the presence of
external periodic driving force (Saif 1999). In the presence of an
external modulation, for example, the spectrum of a tightly binding
potential is a point spectrum (Hawland 1979, 1987, 1989a, 1989b,
Joye 1994).

In the presence of an external modulation, we observe a transition
from point spectrum to a continuum spectrum in the weakly binding
potentials, above a certain critical modulation strength (Delyon
1985, Oliveira 1994, Brenner 1996). Therefore, in the weakly binding
potentials we find localization below a critical modulation
strength, where the system finds a point spectrum.

The quantum delocalization occurs in the Fermi accelerator as a
material wave packet undergoes quantum diffusion for a modulation
strength above $\lambda_u$. In this case the momentum distribution
becomes Maxwellian as we find it in the corresponding classical
case. The presence of the diffusive dynamics above the upper
boundary, shown in figure~\ref{fg:mplocal1}(b), is an illustration
of the quasi-continuum spectrum in the Fermi accelerator above
$\lambda_u$.

\begin{figure}[t]
\begin{center}
\end{center}
\caption{A transition from a localized to a delocalized quantum
mechanical momentum distribution occurs at $\lambda_u=1$, for
$k^{\hspace{-2.1mm}-}=4$. (a) For the modulation strength
$\lambda=0.8<\lambda_u$, we find exponential localization. (b) In
contrast for $\lambda=1.2> \lambda_u$ we find a broad Gaussian
distribution indicating delocalization, similar to Maxwell
distribution in classical diffusion. The initial width of the atomic
wave packet in the Fermi accelerator is, $\Delta z=2$ in position
space, and $\Delta p= k^{\hspace{-2.1mm}-}/2\Delta z=1$ in momentum
space. We take $V=4$ and $\kappa=0.5$. The quantum distributions are
noted after a propagation time $t=3200$.} \label{fg:mplocal1}
\end{figure}


\section{Dynamical recurrences}
\label{dynrev}

The quantum recurrence phenomena are a beautiful combination of
classical mechanics, wave mechanics, and quantum laws. A wave packet
evolves over a short period of time in a potential, initially
following classical mechanics. It spreads while moving along its
classical trajectory, however rebuilds itself after a time called as
classical period. In its long time evolution it follows wave
mechanics and gradually observes a collapse. However, the
discreteness of quantum mechanics leads to the restoration and
restructuring of the wave packet at the time named as quantum
revival time.

In one degree of freedom systems the phenomena of quantum revivals
are well studied both theoretically and experimentally. The quantum
revivals were first investigated in cavity quantum electrodynamics
(Eberly 1980, Narozhny 1981, Yurke 1986). Recently, the existence of
revivals has been observed in atomic (Parker 1986, Alber 1986, Alber
1988, Averbukh 1989, 1989a, Braun 1996, Leichtle 1996, 1996a,
Aronstein 2000) and molecular (Fischer 1995, Vrakking 1996,
Grebenshchikov 1997, Doncheski 2003) wave packet evolution.

The periodically driven quantum systems (Hogg 1982, Haake 1992,
Breslin 1997), and two-degree-of-freedom systems such as stadium
billiard (Tomsovic and Lefebvre 1997) presented the first indication
of the presence of quantum revivals in higher dimensional systems.
Later, it is proved that the quantum revivals are generic to the
periodically driven one degree of freedom quantum systems (Saif
2002, Saif 2005) and to the coupled two degrees-of-freedom quantum
systems (Saif 2005a).


\subsection{Dynamical recurrences in a periodically driven system}

The time evolution of a material wave packet in a one dimensional
system driven by an external periodic field is governed by the
general Hamiltonian, $H$, expressed in the dimensionless form as
\begin{equation}
H=H_0+ \lambda\,V(z)\sin t\;.
\label{eq:sche}
\end{equation}
Here, $H_0$, is the Hamiltonian of the system in the absence of
external driving field. Moreover, $\lambda$ expresses the
dimensionless modulation strength. We may consider $|n\rangle$, and,
$E_n$ as the eigen states and the eigen values of $H_0$,
respectively.

In order to make the measurement of the times of recurrence in the
classical and the quantum domain, we solve the periodically driven
system. In the explicitly time dependent periodically driven system
the energy is no more a constant of motion. Therefore, the
corresponding time dependent Schr\"odinger equation is solved by
using the secular perturbation theory in the region of resonances
(Born 1960).

In this approach, faster frequencies are averaged out and the
dynamical system is effectively reduced to
one degree of freedom. The reduced Hamiltonian is integrable and
its eigen energies and eigen states are considered as the quasi-energy
and the quasi-energy eigen states of the periodically driven system.


\subsubsection{Quasi-energy and quasi-energy eigen states}

In order to solve the time dependent Schr\"odinger equation which
controls the evolution in a periodically driven system, Berman and
Zaslavsky (Berman 1977), and later Flatte' and Holthaus (Flatte'
1996), suggested its solution in an ansatz, as
\begin{equation}
|\psi(t)\rangle=\sum_{n} C_n(t) |n\rangle
\exp\left\{-i\left[E_r+(n-r)\frac{\hbar}{N}\right] \frac{ t}{\hbar
}\right\}\;. \label{eq:1}
\end{equation}
Here, $\hbar$ is the scaled Planck's constant for the driven system,
$E_r$ is the energy corresponding to the mean quantum number $r$,
and $N$ describes the number of resonance.

By following the method of secular perturbation the time dependent
Schr\"odinger equation reduces to
\begin{equation}
i\hbar\frac{\partial g}{\partial t}=
\left[-\frac{{\hbar}^2 N^2 \zeta}{2}\frac{\partial^2}{\partial\theta^2}
-iN\hbar\left(\omega-\frac{1}{N}\right)\frac{\partial}{\partial\theta}
+H_0 +\lambda V\sin(\theta)\right]g(\theta, t)\;.
\label{eq:5}
\end{equation}
The parameters, $\zeta=E''_r/\hbar^2$, and, $\omega=E'_r/\hbar$,
express the nonlinearity in the system and the classical frequency,
respectively. Here, $E''_r$ defines the second derivative of energy
with respect to quantum number $n$ calculated at $n=r$, and $E'_r$
defines the first derivative of the unperturbed energy, calculated
at $n=r$.

The function $g(\theta)$ is related to $C_n(t)$ as,
\begin{eqnarray}
C_n&=&\frac{1}{2\pi}\int\limits_{0}^{2\pi} g(\varphi)
e^{-i(n-r)\varphi}\,d\varphi, \nonumber\\
&=&\frac{1}{2N\pi}\int\limits_{0}^{2N\pi} g(\theta)
e^{-i(n-r)\theta/N}\,d\theta\;. \label{eq:3a}
\end{eqnarray}
We may define $g(\theta, t)$ as
\begin{equation}
g(\theta,t)={\tilde g}(\theta)e^{-i{\cal E}t/\hbar}e^{-i(\omega-1/N)\theta/N\zeta\hbar}
\end{equation}
and take the angle variable $\theta$, as $\theta=2z+\pi/2$. These
substitutions reduce the equation~(\ref{eq:5}) to the standard
Mathieu equation (Abramowitz 1992), which is,
\begin{equation}
\left[\frac{\partial^2}{\partial z^2}
+a-2q\cos(2 z)\right]
\tilde{g}(z)=0\;.
\label{eq:6}
\end{equation}
Here,
\begin{equation}
a=\frac{8}{N^2\zeta\hbar^2}\left({\cal E}-{\bar H}_0+\frac{(\omega-1/N)^2}{2\zeta}\right),
\end{equation}
and $q=4\lambda V/N^2\zeta\hbar^2$.  Moreover, $\tilde{g}_{\nu}(z)$
is a $\pi$-periodic Floquet function, and ${\cal E_{\nu}}$ defines
the quasi-energy of the system for the index $\nu$, given as
\begin{eqnarray}
\nu=\frac{2(n-r)}{N}+\frac{2(\omega-1/N)}{N\zeta\hbar}.
\label{eq:denu}
\end{eqnarray}

Hence, the quasi energy eigen values ${\cal E}_{\nu}$, are defined
as
\begin{eqnarray}
{\cal E}_{\nu}\equiv\frac{\hbar^2N^2\zeta}{8}a_{\nu}(q)-
\frac{(\omega-1/N)^2}{2\zeta}+{\bar H}_0\;. \label{eq:eneg}
\end{eqnarray}
In this way, we obtain an approximate solution for a nonlinear
resonance of the explicitly time-dependent system.


\subsubsection{The dynamical recurrence times}

An initial excitation produced, at $n=r$, observes various time
scales at which it recurs completely or partially during its quantum
evolution. In order to find these time scales at which the
recurrence occur in a quantum mechanical driven system, we employ
the quasi energy, ${\cal E}_{\nu}$, of the system (Saif 2002, Saif
2005).

These time scales $T_{\lambda}^{(j)}$, are inversely proportional to
the frequencies, $\omega^{(j)}$, such that
$T_{\lambda}^{(j)}=2\pi/\omega^{(j)}$, where $j$ is an integer. We
may define the frequency, $\omega^{(j)}$, of the reappearance of an
initial excitation in the dynamical system as
\begin{equation}
\omega^{(j)} =(j!\hbar)^{-1} \frac{\partial^{(j)}{\cal
E}_{\nu}}{\partial n^{(j)} }, \label{omj}
\end{equation}
calculated at $n=r$. The index $j$ describes the order of
differentiation of the quasi energy, ${\cal E}_k$ with respect to
the quantum number, $n$. The equation~(\ref{omj}) indicates that as
value of $j$ increases, we have smaller frequencies which lead to
longer recurrence times.

We substitute the quasi energy, ${\cal E}_k$ as given in
equation~(\ref{eq:eneg}), in the expression for the frequencies
defined in the equation~(\ref{omj}). This leads to the classical
period, $T_{\lambda}^{(1)}=T_{\lambda}^{(cl)}$ for the driven system
as,
\begin{equation}
T_{\lambda}^{(cl)}=[1-M_o^{(cl)}]T_0^{(cl)}\Delta,\label{eq:clt1}
\end{equation}
and, the quantum revival time $T_{\lambda}^{(2)}=T_{\lambda}^{(Q)}$,
as,
\begin{equation}
T_{\lambda}^{(Q)}=[1-M_o^{(Q)}]T_0^{(Q)}.\label{eq:clt2}
\end{equation}
Here, the time scales $T_0^{(cl)}$ and $T_0^{(Q)}$ express the
classical period and the quantum revival time in the {\it absence}
of external modulation, respectively. They are defined as
$T^{(cl)}_0(\equiv 2\pi/\omega)$, and $T^{(Q)}_0 (\equiv
2\pi(\frac{\hbar}{2!}\zeta)^{-1})$. Furthermore,
$\Delta=(1-\omega_N/\omega)^{-1}$, where $\omega_N=1/N$.

The time modification factor $M_o^{(cl)}$ and $M_o^{(Q)}$ are given
as,
\begin{equation}
M_o^{(cl)}=-\frac{1}{2}\left(\frac{\lambda V \zeta\Delta^2} {%
\omega^2} \right)^2 \frac{1}{(1-\mu^2)^2}\, ,
\label{eq:modf1}
\end{equation}
and
\begin{equation}
M_o^{(Q)}=\frac{1}{2}\left(\frac{\lambda V \zeta\Delta^2} {%
\omega^2} \right)^2 \frac{3+\mu^2}{(1-\mu^2)^3}\,,
 \label{eq:modf2}
\end{equation}
where,
\begin{eqnarray}
\mu=\frac{N^2\hbar\zeta\Delta}{2\omega}.
\label{mu}
\end{eqnarray}

Equations (\ref{eq:clt1}) and (\ref{eq:clt2}) express the classical
period and the quantum revival time in a periodically system. These
time scales are function of the strength of the periodic modulation
$\lambda$ and the matrix element, $V$. Moreover, they also depend on
the frequency, $\omega$, and the nonlinearity, $\zeta$, associated
with the {\it un}driven or {\it un}modulated system.

As the modulation term vanishes, that is $\lambda=0$, the
modification terms $M_o^{(cl)}$ and $M_o^{(Q)}$ disappear. Since
there are no resonances for $\lambda=0$, we find $\Delta=1$. Thus,
from equations (\ref{eq:clt1}) and (\ref{eq:clt2}), we deduce that
in this case the classical period and the quantum revival time in
the presence and in the absence of external modulation are equal.

%
\subsection{Classical period and quantum revival time: Interdependence}

The nonlinearity $\zeta$ present in the energy spectrum of the {\it
un}driven system, contributes to the classical period and the
quantum revival time, in the presence and in the absence of an
external modulation. Following we analyze different situations as
the nonlinearity varies.

\subsubsection{Vanishing nonlinearity}
\label{vnn}

In the presence of a vanishingly small nonlinearity in the energy
spectrum, {\it i.e} for $\zeta\approx 0$, the time modification
factors for the classical period $M_o^{(cl)}$ and the quantum
revival time $M_o^{(Q)}$ vanish, as we find in equations
(\ref{eq:modf1}) and (\ref{eq:modf2}). Hence, the modulated linear
system displays recurrence only after a classical period, which is
$T_{\lambda}^{(cl)}=T_0^{(cl)}\Delta=2\pi\Delta/\omega$. The quantum
revival takes place after an infinite time, that is,
$T_{\lambda}^{(Q)}= T_0^{(Q)}=\infty$.


\subsubsection{Weak nonlinearity}

For weakly nonlinear energy spectrum, the classical period,
$T_{\lambda}^{(cl)}$, and the quantum revival time,
$T_{\lambda}^{(Q)}$, in the modulated system are related with,
$T_0^{(cl)}$, and, $T_0^{(Q)}$, of the {\it un}modulated system as,
\begin{equation}
3T_{\lambda}^{(cl)}T_0^{(Q)}+\Delta
T_0^{(cl)}T_{\lambda}^{(Q)}=4\Delta T_0^{(Q)}T_{0}^{(cl)}.
\label{eq:r1}
\end{equation}

As discussed earlier, the quantum revival time, $T_{0}^{(Q)}$,
depends inversely on the nonlinearity, $\zeta$, in the unmodulated
system. As a result $T_{0}^{(Q)}$ and $T_{\lambda}^{(Q)}$, are much
larger than the classical period, $T_{0}^{(cl)}$, and
$T_{\lambda}^{(cl)}$.

The time modification factors $M_o^{(cl)}$ and $M_o^{(Q)}$ are
related as,
\begin{equation}
M_o^{(Q)}=-3 M_o^{(cl)}=3\alpha, \label{moq}
\end{equation}
where
\begin{equation}
\alpha= \frac{1}{2}\left(\frac{\lambda V \zeta
\Delta^2}{{\omega}^2}\right)^2. \label{alphaq}
\end{equation}
Hence, we conclude that the modification factors $M_o^{(Q)}$ and
$M_o^{(cl)}$ are directly proportional to the {\it square} of the
nonlinearity, $\zeta^2$ in the system. From equations (\ref{eq:r1})
and (\ref{moq}) we note that as the quantum revival time
$T_{\lambda}^{(Q)}$ reduces by $3\alpha T_0^{(Q)}$,
$T_{\lambda}^{(cl)}$ increases by $\alpha T_0^{(cl)}$.

In the asymptotic limit, that is for $\zeta$ approaching zero, the
quantum revival time in the modulated and unmodulated system are
equal and infinite, that is $T_{\lambda}^{(Q)}=T_0^{(Q)}=\infty$.
Furthermore, the classical period in the modulated and unmodulated
cases are related as, $T_{\lambda}^{(cl)}=T_0^{(cl)}\Delta$, as
discussed in section~\ref{vnn}.


\subsubsection{Strong nonlinearity}

For relatively strong nonlinearity, the classical period,
$T_{\lambda}^{(cl)}$, and the quantum revival time,
$T_{\lambda}^{(Q)},$ in the modulated system are related with the,
$T_0^{(cl)}$, and, $T_0^{(Q)},$ of the {\it un}modulated system as,
\begin{equation}
T_{\lambda}^{(cl)}T_0^{(Q)}-\Delta T_0^{(cl)}T_{\lambda}^{(Q)}=0.
\label{eq:r2}
\end{equation}
The time modification factors, $M_o^{(Q)}$ and $M_o^{(cl)}$, follow
the relation
\begin{equation}
M_o^{(Q)}=M_o^{(cl)}=-\beta,
\label{relm}
\end{equation}
where
\begin{equation}
\beta= \frac{1}{2}\left(\frac{4\lambda V}{N^2 \zeta \hbar^2}\right)^2.
\label{betaq}
\end{equation}

Hence, for a relatively strong nonlinear case, $\beta$ may approach
to zero. Thus, the time modification factors vanish both in the
classical and the quantum domain. As a result, the equations
(\ref{eq:clt1}) and (\ref{eq:clt2}) reduce to,
$T_{\lambda}^{(cl)}=T_{0}^{(cl)}\Delta$, and
$T_{\lambda}^{(Q)}=T_{0}^{(Q)}$, which proves the equality given in
equation~(\ref{eq:r2}).

The quantity $\beta$, which determines the modification both in the
classical period and in the quantum revival time, is inversely
depending on fourth power of Planck's constant $\hbar$. Hence, for
highly quantum mechanical cases, we find that the recurrence times
remain unchanged.


\section{Dynamical recurrences in the Fermi accelerator}
\label{dynrevf}

For the atomic dynamics in the Fermi accelerator, the classical frequency, $\omega$,
and the nonlinearity, $\zeta$, read as
\begin{equation}
\omega=\left(\frac{\pi^2}{3rk^{\hspace{-2.1mm}-}}\right)^{1/3}
\label{o}
\end{equation}
and
\begin{equation}
\zeta=-\left(\frac{\pi}{9r^2{k^{\hspace{-2.1mm}-}}^2}\right)^{2/3}
\label{z}
\end{equation}
respectively.

In the undriven system, we define the classical period as
$T_0^{(cl)}=2\pi/\omega$, and the quantum revival time as
$T_0^{(Q)}=2\pi/k^{\hspace{-2.1mm}-}\zeta$. Hence, we find an
increase in the classical period and in the quantum revival time of
the undriven Fermi accelerator as the mean quantum number $r$
increases.

The calculation for the matrix element, $V$, for large $N$, yields
the result,
\begin{equation}
V\cong -\frac{2E_0}{N^2\pi^2},
\end{equation}
and the resonance, $N$, takes the value
\begin{equation}
N=\frac{\sqrt{2E_N}}{\pi}.
\end{equation}

The time required to sweep one classical period and the time
required for a quantum revival are modified in the driven system. In
the classical domain and in the quantum domain the modification term
becomes,
\begin{equation}
M_o^{(cl)}=\frac{1}{8}\left(\frac{\lambda } {E_N} \right)^2
\frac{1}{(1-\mu^2)^2}\, , \label{eq:modfe1}
\end{equation}
and
\begin{equation}
M_o^{(Q)}=\frac{1}{8}\left(\frac{\lambda } {E_N} \right)^2 \frac{3+\mu^2}{(1-\mu^2)^3}%
\,  \label{eq:modfe2}
\end{equation}
respectively, where
\begin{equation}
\mu=-\frac{k^{\hspace{-2.1mm}-}}{4}\frac{1}{E_0}\sqrt{\frac{E_N}{E_0}}.
\end{equation}
Therefore, the present approach is valid for a smaller value of the
modulation strength, $\lambda$ and a larger value of the resonance,
$N$.

A comparison between the quantum revival time, calculated with the
help of equation (\ref{eq:modfe2}),
with those obtained numerically for an atomic wave-packet bouncing
in a Fermi accelerator shows a good agreement, and thus supports
theoretical predictions (Saif 2000e, 2002).

The quantum recurrences change drastically at the onset of a
periodic modulation. However the change depends upon the initial
location of the wave packet in the phase space. In order to emphasis
the fascinating feature of quantum chaos, we note that for an
initially propagated atomic wave packet in the {\it un}driven Fermi
accelerator the quantum revivals exist. However, they disappear
completely in the presence of modulation if the atom originates from
the center of a primary resonance. In the present situation the
atomic wave packet reappears after a classical period only (Saif
2005a), as shown in figure \ref{fg:rev34}.

\begin{figure}[t]
\begin{center}
\end{center}
\caption{The change in revival phenomena for the wave packet
originating from the center of a resonance: (above) The square of
auto-correlation function, $C^2=|<\psi(0)|\psi(t)>|^2$, is plotted
as a function of time, $t$, for an atomic wave packet. The wave
packet initially propagates from the center of a resonance in atom
optics Fermi Accelerator and its mean position and mean momentum are
$z_0=14.5$ and $p_0=1.45$, respectively. Thick line corresponds to
numerically obtained result for the Fermi accelerator and dashed
line indicates quantum revivals for a wave packet in harmonic
oscillator. The classical period is calculated to be $4\pi$ for
$\lambda=0.3$. (bellow) The square of the auto-correlation function
as a function of time for $\lambda=0$. As the modulation strength
$\lambda$ vanishes the evolution of the material wave packet changes
completely. The wave packet experiences collapse after many
classical periods and, later, displays quantum revivals at
$T_0^{(Q)}$. The inset displays the short time evolution of the wave
packet comprising many classical periods for in the absence of
external modulation. (Saif 2005a).} \label{fg:rev34}
\end{figure}

We may understand this interesting property as we note that at a
resonance the dynamics is effectively described by an oscillator
Hamiltonian (Saif 2000c),
\begin{equation}
H= -\frac{\partial^2}{\partial\varphi^2}+ V_0\cos\varphi.
\end{equation}
Therefore, when $\varphi\ll 1$, the Hamiltonian reduces to the
Hamiltonion for a harmonic oscillator, given as
\begin{equation}
H\approx -\frac{\partial^2}{\partial\varphi^2}-
\frac{V_0}{2}\varphi^2.
\end{equation}
The Hamiltonian controls the evolution of an atomic wave packet
placed at the center of a resonance. This analogy provides us an
evidence that if an atomic wave packet is placed initially at the
center of a resonance, it will always observe revivals after each
classical period only, as in case of a harmonic oscillator.

In addition, this analogy provides information about the level
spacing and level dynamics at the center of a resonance in the Fermi
accelerator as well. Since in case of harmonic oscillator the
spacing between successive levels is always equal, we conclude that
the spacing between quasi-energy levels is equal at the center of
resonance in a periodically driven system.

For a detailed study on the effect of initial conditions on the
existence of the quantum recurrences, the information about the
underlying Floquet quasi-energies, and the use of quantum
recurrences to understand different dynamical regimes we refer to
the references (Saif 2000c, 2005, 2005a).

%
\section{Chaos in complex atom optics systems}
\label{sum}

At the dawn of the 20$^{th}$ century, Max Planck provided the
conceptual understanding of the radiation fields as particles. The
idea of the wave nature for material particles, coined by De
Broglie, thus gave birth to the wave-particle duality principle.

In later years the wave-particle duality became a prime focus of
interest in the electron optics and neutron optics. However, easily
controllable momenta and rich internal structure of atoms, provided
inherent advantages to atom optics over electron optics and neutron
optics.

The enormous work of last three decades in atom optics has opened up
newer avenues of research. These include many theoretical and
experimental wonders, such as cooling of atoms to micro kelvin and
nano kelvin scales, Bose Einstein condensation, and atom lasers.

The study of the classical and the quantum chaos in atom optics
started after a proposal by Graham, Shloutman, and Zoller. They
suggested that an atom passing through a phase modulated standing
light wave may exhibit chaos as the atom experiences a random
transfer of momentum.

Later, numerous periodically driven systems were studied in the
frame work of atom optics to explore characteristics of the
classical and the quantum chaos. On the one hand, study of these
systems enabled the researchers to understand the theory of chaos in
laboratory experiments, and on the other hand has brought to light
newer generic properties, such as quantum dynamical recurrence
phenomena in chaotic systems.

\subsection{ Atom in a periodically modulated standing wave field}

In the original Graham Schloutman and Zoller (GSZ) proposal, an
altra cold atom in a phase modulated standing wave field exhibits
Anderson like localization phenomenon in the momentum space. The
exponential localization is visible as the momentum distribution
exhibits exponential decay in the stochastic region.

There exist domains on the modulation strength, where the GSZ system
is almost integrable and the classical dynamics is regular. The
exponential localization of the momentum distribution vanishes for
these modulation strengths and occurs otherwise.
Thus, the dynamical localization occurs over windows on modulation
strength, separated by the domains which support integrable
dynamics.


The atom in the phase modulated standing wave has one more
interesting feature. The classical phase space is a mixed phase
space containing islands of stability in a stochastic sea. The
modulation amplitude determines the ratio between the phase space
areas corresponding to stability and chaos. This ratio serves as a
measure of chaos.

These dynamical features are experimentally verified by the group of
Mark Raizen at Texas (Raizen 1999). In the experiment, sodium atoms
were stored and laser-cooled in a magneto-optical trap. After
turning off the trapping fields and switching on a standing light
wave which has phase modulation the atoms experience momentum kicks.
The experiment demonstrates the exponential localization of the
atoms in the momentum space (Latka 1995).

The localization windows of GSZ model have a remarkable difference
from the localization window of the Fermi accelerator, discussed in
section~\ref{sec:locwin}. In GSZ model, outside the localization
windows, the dynamics is stable and bounded both classically and
quantum mechanically.
In contrast, in the Fermi accelerator the dynamics is bounded for
modulation strengths below the localization window, whereas it is
unbounded and chaotic for modulation strengths above the upper bound
of the localization window both in the classical and quantum domain.

\subsection{Delta kicked rotor}

A rotor is a particle connected by a massless rod to a fixed
point. While the particle rotates frictionless around this point it
experiences an instantaneous periodic force.
The kicked rotor model has extensively been studied in classical chaos
and at the same time it has become the beauty of quantum chaos.


Historically the system provided the first study of dynamical
localization in a periodically driven time dependent system (Casati
1979). Later, a formal connection was established between the kicked
rotor model and one dimension tight binding Anderson model with a
time-dependent pseudo-random potential (Fishman 1982, Fishman 1984).
This led to recognition that the quantum suppression of classically
chaotic diffusion of momentum of rotor is a dynamical version of the
Anderson localization.

The work of Mark Raizen, as discussed in section~\ref{krm}, built a
bridge between atom optics and delta kicked rotor model (Raizen
1999). The work has led to the experimental verification of
dynamical localization in delta kicked rotor (Moore 1995), to study
the classical anomalous diffusion in quantum domain (Klappauf 1998),
and to develop newer methods for cooling atoms (Ammann 1997).


%

\subsection{Ion in a Paul trap}

The idea of extension of two dimensional focusing of charged and
neutral particles to three dimensions, led the invention of
sophisticated traps and won the inventor, Wolfgang Paul, Nobel prize
in 1989 (Paul 1990). A full quantum treatment of an ion in a Paul
trap has been given in reference (Glauber 1992, Stenholm 1992).

The trapped ions have been considered to generate the non classical
states of motion of ions (Meekhof 1996, Monroe 1996, Cirac 1996),
quantum logic gates (Monroe 1995, Cirac 1995), tomography (Leibfried
1996, Wallentowitz 1995, Poyatos 1996, D'Helon 1996) and endoscopy
(Bardroff 1996) of the density matrix .

Two or more than two ions in a Paul trap may display chaos due to
ion-ion Coulomb repulsion (Hoffnagle et al. 1988, Bl\"umel et al.
1989, Brewer et al. 1990). However a single ion moving in a trap in
the presence of a standing light field, displays chaos as well
(Chac\'on and Cirac 1995, Ghaffar et al. 1997). The effective
Hamiltonian of the system becomes
\begin{equation}
H=\frac{p^2}{2}+ \frac{1}{2}W(t)x^2 + V_0\cos(x+2\phi),
\end{equation}
where $W(t)=a +2q\cos t$ and $V_0$ is the effective coupling
constant. The parameters $a$ and $q$ denote the dc and ac voltage
applied to the trap (Paul 1990).

The ion trap has been proposed to study the dynamical localization
both in the position space and in the momentum space (Ghafar 1997,
Kim 1999). In addition, the quantum revivals have also been
predicted in the system (Breslin 1997).

\subsection{Fermi-Ulam Accelerator}

In the Fermi-Ulam accelerator model a particle bounces off a
periodically vibrating wall and returns to it after observing
reflection from a static wall, parallel to the vibrating wall. Based
on the original idea of Fermi (Fermi 1949), the classical Fermi-Ulam
accelerator has been investigated extensively (Lichtenberg 1983,
1992). The system exhibits chaotic, transitional and periodic
solutions (Zaslavskii 1964, Brahic 1971, Lieberman 1972, Lichtenberg
1980).

The classical dynamics of the Fermi-Ulam accelerator model is
expressible by a highly nonlinear mapping which connects the phase
space position of the system at $n$th collision with the ($n+1$)th
collision. The study of the mapping revels that the orbit of a
particle in the system can be stochastic (Zaslavskii 1964). However,
the energy of the particle remains bounded and the unlimited
acceleration as proposed by Enrico Fermi is absent (Zaslavskii 1964,
Pustilnikov 1983, 1987).

In 1986, the quantum Fermi-Ulam accelerator was introduced (Jos\'e
1986, Reichl 1986) which was later studied extensively to explore
the quantum characteristics (Visscher 1987, Jose 1991, Scheininger
1991, Chu 1992). The beautiful numerical work of Jos\'e and Cordery
has demonstrated a clear transition from Poisson statistics of
quasienergies in the accelerator system to Wigner distribution
(Jos\'e 1986). The Poisson statistics implies integrability and
Wigner distribution corresponds to nonintegrability (Brody 1981),
hence, the transition is related to a transition from regular to
irregular behavior.

The study of quasi-energy spectrum of the quantum Fermi-Ulam system
has led to establish that in general the time evolution of a quantum
particle in the system is recurrent, and thus the energy growth
remains bounded. However, for some particular wall oscillations the
particle may gain unbounded acceleration (Seba 1990).

\section{Dynamical effects and Decoherence}
\label{dedecoh}

Quantum interference effects are extremely fragile (Giulini 1996),
and sensitive to dissipation and decoherence
(Walls 1994, Zurek 1991).
The phenomenon of dynamical localization is an interference effect.
It is therefore extremely susceptible to noise and decoherence
(Zurek 1995).

First experiments (Bl\"umel 1996) using the dynamical localization
of Rydberg atoms in microwave fields have clearly verified the
destructive nature of noise. In atom optics the experimental
realization of the kicked rotor has opened up a new door to study
the influence of noise and dissipation on such quantum interference
effects (Klappauf 1998).

The presence of amplitude noise in kicked rotor destroys the
exponential localization and broad shoulders develop in the momentum
distributions. Furthermore, by tuning the laser field close to the
resonance of the atom, the atom gets a non zero transition
probability to jump to the excited state, and to spontaneously decay
to the ground state. This spontaneous emission event destroys the
coherence and thus the localization, as shown in figure~\ref{decoh}.

\begin{figure}[t]
\begin{center}
\end{center}
\caption{Comparison of the momentum distribution is made in the
kicked rotor for the cases of (a) no noise, (b) $62.5\%$ amplitude
noise, and (c) dissipation from 13\%/kick spontaneous scattering
probability. Time steps shown are zero kicks (light solid line), 17
kicks (dashed-dotted), 34 kicks (dashed), 51 kicks (dotted), and 58
kicks heavy solid for (a) and (c), and 0, 16, 32, 52, and 68 kicks,
respectively, for (b). (Klappauf {\it et al.} 1998).} \label{decoh}
\end{figure}

For more experiments on decoherence in dynamical localization we
refer to (Ammann 1997). Moreover, there exists an extensive
literature on the influence of spontaneous emission on dynamical
localization, see for example (Graham 1996, Goetsch 1996, Dyrting
1996, Riedel 1999).

\section{Acknowledgment}

It is a pleasure to acknowledge the consistent check on the
completion of the report by Prof. Eichler. The task became less
daunting in the presence of his suggestions and moral support.

During the early stages of the report, I have enjoyed numerous
discussions with W. P. Schleich, I.~Bialynicki-Birula and M.
Fortunato. Indeed I thank them for their time and advice. In the
last few years I benefited from the knowledge and patience of many
of my friends. Indeed I thank all of them, in particular I mention
G. Alber, I.~Sh.~Averbukh, P. Bardroff, V. Balykin, G. Casati, H.
Cerdeira, A. Cronin, J. Diettrich, J. Dalibard, M. Fleischhauer, M.
Al-Ghafar, R. Grimm, I. Guarneri, F. Haake, K. Hakuta, M. Holthaus,
H. Hoorani, Fam Le Kien, I. Marzoli, P. Meystre, V. Man'ko, B.
Mirbach, G. J. Milburn, M. M. Nieto, A. Qadir, V. Savichev, R.
Schinke, P. Seba, D. Shepelyansky, F. Steiner, S. Watanabe, P.
T\"orm\"a, V. P. Yakovlev and M. S. Zubairy. I submit my thanks to
K. Blankenburg, A. Qadir and K. Sabeeh for a careful study of the
manuscript.

I am thankful to the Universit\"at Ulm, Germany and to the National
Center for Physics, Pakistan for the provision of the necessary
computational facilities. I am also grateful to the Abdus Salam
International Center for Theoretical Physics, Trieste, Italy to
provide support conducive to complete the project.

The author is partially supported by the Higher Education
Commission, Pakistan through the research grant R\&D/03/143, and
Quaid-i-Azam University research grants.


\begin{thebibliography}{999}

\bibitem{kn:abra}  Abramowitz M. and I.A. Stegun, {\it Handbook of
Mathematical Functions},  (Dover, New York, 1992).

\bibitem{Adams} Adams C.~S., M.~Sigel and J.~Mlynek, Phys. Rep. {\bf 240}
         (1994) 143.

\bibitem{kn:alber} Alber G., H. Ritsch, and P. Zoller, Phys. Rev. A {\bf 34}%
(1986) 1058.

\bibitem{kn:alber1}  Alber G. and P. Zoller, Phys. Rev. A {\bf 37}
(1988) 377.

\bibitem{kn:alber2}  Alber G. and P. Zoller, Phys. Rep. {\bf 199},
(1990) 231.

\bibitem{Altland 1996} Altland A., and M. R. Zirnbauer, Phys. Rev.
Lett. {\bf 77} (1996) 4536.

\bibitem{kn:amino}  Aminoff C.~G., A.~M.~Steane, P.~Bouyer,
P.~Desbiolles, J.~Dalibard, and C.~Cohen-Tannoudji, Phys. Rev. Lett.
{\bf 71} (1993) 3083.

\bibitem{kn:amma}  Ammann H. and N.~Christensen, Phys. Rev. Lett. {\bf
78} (1997) 2088.

\bibitem{Cornell} Anderson M.~H., J.~R. Ensher, M.~R. Matthews, C.~E. Wiemann,
and E.~A. Cornell, Science {\bf 269} (1995) 198.

\bibitem{Anderson} Anderson P.~W., Phys. Rev. {\bf 109} (1958) 1492.

\bibitem{Anderson1}Anderson P.~W., Phys. Rev. {\bf 115} (1959) 2.

\bibitem{kn:andre} Andrews M., Am. J. Phys. {\bf 66} (1998) 252.

\bibitem{kn:arim}  Arimondo E. and H.~A.~Bachor, (eds.) {\it Special
issue
on Atom Optics}, J. Quant. Semicl. Opt. {\bf 8} (1996) 495.

\bibitem{kn:arnd}  Arndt M., A.~Buchleitner, R.~N.~Mantegna, and
H.~Walther,
Phys. Rev. Lett. {\bf 67} (1991) 2435.

\bibitem{kn:arn}  Arnol'd V. I, ed., {\it Dynamical Systems III.
Encyclopedia of Mathematical Sciences}, Vol. 3 (Springer, Berlin, 1988).

\bibitem{kn:arav}  Arnol'd V. I. and A.~Avez, {\it Ergodic Problems of
Classical Mechanics} (Benjamin, New York, 1968).

\bibitem{kn:stroud}  Aronstein D. L. and C.~R.~Stroud, Jr.,  Phys. Rev. A
{\bf 62} (2000) 022102.

\bibitem{kn:artuso1}  Artuso R., Physica D {\bf 109} (1997)a 1.

\bibitem{kn:artuso2}  Artuso R., Phys. Rev. E {\bf 55} (1997)b 6384.

\bibitem{kn:aver}  Averbukh I. Sh. and N.F. Perel'man, Phys. Lett. A {\bf 139%
} (1989) 449.

\bibitem{kn:aper}  Averbukh I. Sh. and N.~F.~Perel'man, Sov. Phys. JETP
{\bf %
69} (1989)a 464.

\bibitem{Averbuckh 1995} Averbuckh V., N. Moiseyev, B. Mirbach, and H. J. Korsch, Z. Phys. D
{\bf 35} (1995) 247.










\bibitem{kn:badri}  For a discussion of a classical and
quantum-mechanical two-dimensional Fermi accelerating disk, see:
Badrinarayanan R., J.~V.~Jos\'{e}, and G.~Chu, Physica D {\bf 83}
(1995) 1.

\bibitem{kn:baly}  The first experimental demonstration of atoms
bouncing on an atomic mirror was made by Balykin V. I.,
V.~S.~Letokhov, Y.~B.~Ovchinnikov, and A.~I.~Sidorov, Pis'ma Zh.
Eksp. Teor. Fiz. {\bf 45} (1987) 282 [JETP Lett. {\bf 45} (1987)
353].

\bibitem{kn:baly2}  Balykin V. I., V.~S.~Letokhov, Y.~B.~Ovchinnikov,
and
A.~I.~Sidorov, Phys. Rev. Lett. {\bf 60} (1988) 2137.

\bibitem{kn:baly3}  Balykin V. I. and V.~S.~Letokhov, Appl. Phys. B
{\bf 48}
(1989) 517.

\bibitem{balyhak} Balykin V. I., K. Hakuta, Fam Le Kien, J. Q. Liang, and M. Morinaga,
submitted to Phys. Rev. A. (2004)

\bibitem{kn:bard}  Bardroff P. J., I.~Bialynicki-Birula,
D.~S.~Kr\"{a}hmer,
G.~Kurizki, E.~Mayr, P.~Stifter, and W.~P.~Schleich, Phys. Rev. Lett.
{\bf 74%
} (1995) 3959.

\bibitem{Bardroff 1996} Bardroff P. J., C. Leichtle, G. Schrade, and W. P. Schleich,
Phys. Rev. Lett. {\bf 77} (1996) 2198.

\bibitem{Bayfield} Bayfield J. E., and P. M. Koch, Phys. Rev. Lett. {\bf 33} (1974) 258.

\bibitem{kn:bayf}  Bayfield J. E., G.~Casati, I.~Guarneri, and
D.~W.~Sokol,
Phys. Rev. Lett. {\bf 63} (1989) 364.

\bibitem{kn:becas}  Benvenuto F., G.~Casati, and D.~L.~Shepelyansky,
Phys.
Rev. A {\bf 55} (1997) 1732.

\bibitem{Benvenuto 1991}  Benvenuto F., G.~Casati, I.~Guarneri, and
D.~L.~Shepelyansky, Z. Phys. B {\bf 84} (1991) 159.

\bibitem{kn:berk}  Berkhout J. J., O.~J.~Luiten, I.~D.~Setija,
T.~W.~Hijmans, T.~Mizusaki, J.~T.~M.~Walraven, Phys. Rev. Lett. {\bf
63} (1989) 1689.

\bibitem{Berman1} Berman G. P., and G. M. Zaslavsky, Phys. Lett. A {\bf 61} (1977) 295.

\bibitem{BermanZaslavsky} Berman G. P., and A. R. Kolovsky, Phys. Lett. A {\bf 95} (1983) 15.

\bibitem{berry} Berry M. V., in {\it Cahotic Behavior of Deterministic Systems}, ed. G. Iooss,
R. H. G. Helleman, and R. Stora, Les Houches Summer School
Proceedings {36}, (North-Holland, Amsterdam, 1981).

\bibitem{Blumel 1989} Bl\"umel R., R. Graham, L. Sirko, U. Smilansky, H. Walther and
K. Yamada, Phys. Rev. Lett. {\bf 62} (1989) 341.

\bibitem{Blumel} Bl\"umel R., A.~Buchleitner, R.~Graham, L.~Sirko, U.~Smilansky, H.~Walther,
Phys. Rev. A {\bf 44} (1996) 4521.

\bibitem{Blumel 1989b} Bl\"umel R., C. Kappler, W. Quint, and H.
Walther, Phys. Rev. A {\bf 40} (1989) 808.

\bibitem{Blumel 1997} Bl\"umel R. and W. P. Reinhardt, {\it  Chaos in atomic Physics},
(Cambridge, New York, 1997).

\bibitem{Bohigas} Bohigas O., M. J. Giannoni, and C. Schmit, Phys. Rev. Lett. {\bf 52} (1984) 1.

\bibitem{kn:bord}  Bordo V. G., C.~Henkel, A.~Lindinger, and
H.~G.~Rubahn,
Opt. Commun. {\bf 137} (1997) 249.

\bibitem{kn:born}  Born M., {\it Mechanics of the atom} (Ungar, New York,
1960).

\bibitem{Hulet} Bradley C.C., C.~A. Sackett, J.~J. Tollett, and R.~G. Hulet,
Phys. Rev. Lett. {\bf 75} (1995) 1687.

\bibitem{Brahic} Brahic A., J. Astrophys., {\bf 12} (1971) 98.

\bibitem{kn:brau}  Braun P. A. and V.~I.~Savichev,  J. Phys. B {\bf 29}
(1996) L329.

\bibitem{Brenner 1996}  Brenner N. and S.~Fishman, Phys. Rev. Lett. {\bf
77}
(1996) 3763.

\bibitem{kn:bres}  Breslin J. K., C.A. Holmes, and G.J. Milburn,  Phys. Rev.
A {\bf 56} (1997) 3022.

\bibitem{breuer1} Breuer H. P., K. Dietz, M. Holthaus,
J. Phys. B {\bf 22} (1989) 3187.

\bibitem{Breuer 1991a} Breuer H. P. and M. Holthaus,
Ann. Phys. NY {\bf 211} (1991)a 249.

\bibitem{Breuer 1991b} Breuer H. P., K. Dietz, M. Holthaus,
J. Phys. B {\bf 24} (1991)b 1343.

\bibitem{Brivio 1988} Brivio G. P., G. Casati, L. Perotti, and I.
Guarneri, Physica D, {\bf 33} (1988) 51.

\bibitem{Brewer 1990} Brewer R. G., J. Hoffnagle, R. G. DeVoe, L. Reyna, and
W. Henshaw, Nature {\bf 344} (1990) 305.

\bibitem{Brody} Brody T. A., J. Flores, J.B. French, A. Mello, A. Pandy,
and S. S. M. Wong, Rev. Mod. Phys. {\bf 53} (1981) 385.











\bibitem{Casati1} Casati G., B.~V.~Chirikov, J.~Ford, and F.~M.~Izrailev,
               Lecture Notes Phys., {\bf 93} (1979) 334.

\bibitem{Casati2} Casati G., B.~V.~Chirikov, and D.~L.~Shepelyansky,
               Phys. Rev. Lett. {\bf 53} (1984) 2525.

\bibitem{CasatiC} Casati G., B.~V.~Chirikov, I.~Guarneri, and D.~L.~Shepelyansky,
               Phys. Rep. {\bf 154} (1987) 77.

\bibitem{CasatiP1} Casati G., and T. Prosen, Phys. Rev. E {\bf 59} (1999)a R2516.

\bibitem{CasatiP2} Casati G., and T. Prosen, Physica D {\bf 131} (1999)b 293.

\bibitem{casati1999c} Casati G., G. Maspero, and D.~Shepelyansky, Phys. Rev. Lett.
{\bf 82} (1999)c 524.

\bibitem{Chacon 1995} Chac\'on R., and J. I. Cirac, Phys. Rev. A {\bf 51} (1995) 4900.

\bibitem{kn:chen}  Chen Wen-Yu and G.~J.~Milburn, Phys. Rev. E {\bf
56} (1997) 351.

\bibitem{kn:chen1} Chen Wen-Yu and G.~J.~Milburn, Phys. Rev. A {\bf
51}
(1995) 2328.

\bibitem{Chirikov 1979}  Chirikov B. V., Phys. Rep. {\bf 52} (1979) 263.

\bibitem{chirikov1984}  Chirikov B. V. and D. L.~Shepelyansky, Physica (Amsterdam)
{\bf 13D} (1984) 395.

\bibitem{Chirikov 1986}  Chirikov B. V. and D. L.~Shepelyansky, Radiofizika
{\bf 29} (1986) 1041.

\bibitem{chirikov1988}  Chirikov B. V. and D. L.~Shepelyansky, in {\it Renormalization Group}, edited by
D. V. Shirkov, D. I. Kazakov, and A. A. Vladimirov (World Scientific, Singapore, 1988), p.221.

\bibitem{chirikov1999}  Chirikov B. V. and D. L.~Shepelyansky, Phys. Rev. Lett.
{\bf 82} (1999) 528.


\bibitem{kn:chri}  Christ M., A.~Scholz, M.~Schiffer, R.~Deutschmann,
and W.~Ertmer, Opt. Commun. {\bf 107} (1994) 211.

\bibitem{Chu 1992} Chu G., and J. V. Jos\'e, J. Stat. Phys. {\bf 68} (1992) 153.

\bibitem{Cirac 1996} Cirac J. I. et al, Adv. At. Mol. Phys. {\bf 37} (1996) 237.

\bibitem{Cirac 1995} Cirac J. I., and P. Zoller, Phys. Rev. Lett. {\bf 74} (1995) 4091.

\bibitem{kn:cohen}  Cohen-Tannoudji C., B.~Diu, and F.~Lalo\"{e}, {\it
Quantum Mechanics} (John Wiley, New York, 1977).

\bibitem{kn:cook}  The possibility of using an evanescent wave as an
optical mirror for atoms was first proposed by Cook R. J. and
R.~K.~Hill, Opt. Commun. {\bf 43} (1982) 258.












\bibitem{Dana 1995} Dana I., E. Eisenberg, and N. Shnerb, Phys. Rev.
Lett. {\bf 74} 686.

\bibitem{Ketterle} Davis K. B., M.~-O. Mewes, M.~R. Andrews, N.~J. van Druten,
D.~S. Durfee, D.~M. Kurn, and W.~Ketterle, Phys. Rev. Lett. {\bf 75} (1995) 3969.

\bibitem{delyon} Delyon F., B. Simon, B. Souillard, Ann. Inst.
                    H. Poincar\'e {\bf 42} (1985) 283.


\bibitem{Dembinski 1995} Dembi\'nski S. T., A. J. Makowski, P. Peplowski,
                 J. Phys. A: Math. Gen. {\bf 28} (1995) 1449.

\bibitem{kn:desb}  Desbiolles P. and J.~Dalibard, Opt. Commun. {\bf
132} (1996) 540.

\bibitem{kn:desk}  Desko R. D. and D.~J.~Bord, Am. J. Phys. {\bf 51}
(1983)
82.

\bibitem{D'Helon 1996} D'Helon C., and G. J. Milburn,  Phys. Rev. A {\bf 54} (1996) R25.

\bibitem{kn:donch} Doncheski M. A., and R. W. Robinett, arXiv:quant-ph/0307079.

\bibitem{kn:dow}  Dowling J. P. and J.~Gea-Banacloche, Adv. At. Mol.
Opt. Phys. {\bf 37} (1996) 1.

\bibitem{Dyrting} Dyrting S., and G.~J.~Milburn, Phys. Rev. A {\bf 51} (1996) 3136.








\bibitem{kn:eber}  Eberly J. H., N. B. Narozhny, and J. J. SanchezMondragon,
Phys. Rev. Lett. {\bf 44}, 1323 (1980).

\bibitem{Esslinger 1993}  Esslinger T., M.~Weidem\"{u}ller, A.~Hemmerich, and
T.~W.~H\"{a}nsch, Opt. Lett. {\bf 18} (1993) 450.










\bibitem{kn:fermi}  Fermi E., Phys. Rev. {\bf 75} (1949) 1169.

\bibitem{kn:feron} Feron S., J. Reinhardt, S. Le Boiteux, O. Gorceix, J. Baudon,
M. Ducloy, J. Robert, Ch. Miniatura, S. Nic Chormaic, H. Haberland, and V. Lorent,
Opt. Comm. {\bf 102} (1993) 83.

\bibitem{kn:fisc}  Fischer I., D. M. Villeneuve, M. J. J. Vrakking, and A.
Stolow, J. Chem Phys. {\bf 102}, 5566 (1995).

\bibitem{kn:fish}  Fishman S., D.~R.~Grempel, and R.~E.~Prange, Phys.
Rev.
Lett. {\bf 49} (1982) 509.


\bibitem{Fishman 1984}  Fishman S., D.~R.~Grempel, and R.~E.~Prange, Phys.
Rev. A {\bf 29} (1984) 1639.

\bibitem{kn:flat}  Flatte' M. E. and M. Holthaus, Ann. Phys. NY {\bf 245}
(1996) 113.

\bibitem{kn:flei}  Fleischhauer M. and W. P. Schleich, Phys. Rev. A {\bf 47}%
, (1993) 4258.











\bibitem{kn:galv}  Galvez E. J., B.~E.~Sauer, L.~Moorman, P.~M.~Koch,
and
D.~Richards, Phys. Rev. Lett. {\bf 61} (1988) 2011.

\bibitem{kn:gea}  Gea-Banacloche J., Am. J. Phys. {\bf 67} (1999) 776.

\bibitem{kn:moham}  Ghafar M. El, P.~T\"{o}rm\"{a}, V.~Savichev,
E.~Mayr,
A.~Zeiler, and W.~P.~Schleich, Phys. Rev. Lett. {\bf 78} (1997) 4181.

\bibitem{kn:gibb}  Gibbs R. L., Am. J. Phys. {\bf 43} (1975) 25.

\bibitem{kn:Joos} See for example Giulini D., E.~Joos, et al.
                  {\it Decoherence and the Appearance of a
                  Classical World in Quantum Theory}
                  (Springer, Berlin, 1996)

\bibitem{Glauber 1992} Glauber R. J., in Foundations of Qunatum Mechanics,
           edited by T. D. Black,
           M. M. Nieto, H. S. Pilloff, M. O. Scully, and R. M. Sinclair
           (World Scientific, Singapore, 1992).

\bibitem{Goetsch} Goetsch P. and R.~Graham, Phys. Rev. A {\bf 54} (1996) 5345.

\bibitem{kn:good}  Goodins D. A. and T.~Szeredi, Am. J. Phys. {\bf 59}
(1991) 924.

\bibitem{Gorin 1997} Gorin T., H. J. Korsch and B. Mirbach, Chem Phys. {\bf 217} (1997) 147.

\bibitem{kn:grah}  Graham R., M.~Schlautmann, and P.~Zoller, Phys.
Rev. A{\bf 45} (1992) R19.

\bibitem{Graham1} Graham R., and S.~Miyazaki, Phys. Rev. A {\bf 53} (1996) 2683.

\bibitem{kn:greb}  Grebenshchikov S. Yu., C. Beck, H. Fl\"othmann, D. H.
Mordaunt and  R. Schinke, Chem. Phys. Lett. {\bf 271} (1997) 197.

\bibitem{kn:gree}  Greene J., J. Math. Phys. {\bf 20} (1979) 1183.

\bibitem{kn:carp}  Grossmann F., J.-M.~Rost, and W.~P.~Schleich,
J. Phys. A: Math. Gen. {\bf 30} (1997) L277.

\bibitem{Guarneri} Guarneri I., and F. Borgonovi, J. Phys. A: Math. Gen. {\bf 26} (1993) 119.

\bibitem{kn:gutz}  Gutzwiller M. C., {\it Chaos in Classical and
Quantum Mechanics} (Springer, Berlin, 1992).












\bibitem{kn:haak}  Haake F., {\it Quantum signatures of Chaos}
(Springer,
Berlin, 1992).

\bibitem{Haake 2001}  Haake F., {\it Quantum signatures of Chaos}
(Springer,
Berlin, 2001).

\bibitem{kn:hamm} Hammes M., D. Rychtarik, B. Engeser, H.-C. Nägerl, and R. Grimm
Phys. Rev. Lett. {\bf 90} (2003) 173001.

\bibitem{hawland1} Hawland J., Indiana Univ. Math. J. {\bf 28} (1979) 471.

\bibitem{hawland2} Hawland J., J. Funct. Anal. {\bf 74} (1987) 52.

\bibitem{hawland3} Hawland J., Ann. Inst. H. Poincare' {\bf 49} (1989)a 309.

\bibitem{hawland4} Hawland J., Ann. Inst. H. Poincare' {\bf 49} (1989)b 325.


\bibitem{kn:kaiser}  Henkel C., A.~M.~Steane, R.~Kaiser, and
J.~Dalibard, J. Phys. II (France) {\bf 4} (1994) 1877.

\bibitem{Henkel} Henkel C., K. M{\o}lmer, R. Kaiser, N. Vansteenkiste, C. Westbrook, A. Aspect,
            Phys. Rev. A {\bf 55} (1997) 1160.

\bibitem{kn:hogg}  Hogg T. and B. A. Huberman, Phys. Rev. Lett. {\bf 48} (1982) 711.

\bibitem{Hoffnagle 1988} Hoffnagle J., R. G. DeVoe, L. Reyna, and R. G. Brewer,
Phys. Rev. Lett. {\bf 61} (1988) 255.

\bibitem{kn:hinds1}  Hughes I. G., P.~A. Barton, M.~G. Boshier and
E.~A.~Hinds,
J. Phys. B {\bf 30} (1997) 647.

\bibitem{kn:hinds2}  Hughes I. G., P.~A. Barton, and E.~A.~Hinds, J. Phys.
B {\bf 30} (1997) 2119.










\bibitem{Iqbal 2005} Iqbal S., Qurat ul Ann, and F. Saif, in press.

\bibitem{Izrailev1} Izrailev F. M., Phys. Rep. {\bf 196} (1990) 299.







\bibitem{kn:jack}  Jackson J. D., {\it Classical Electrodynamics,
}(John Wiley, New York, 1965).

\bibitem{Jose} Jos{\'e} J. V., and R. Cordery, Phys. Rev. Lett. {\bf 56} (1986) 290.

\bibitem{kn:jos}  Jos\'{e} J. V., in {\it Quantum Chaos}, Proceedings
of the Adriatico Research Conference and Miniworkshop, Trieste,
Italy, 4 June--6 July 1990, edited by H.~A.~Cerdeira, R.~Ramaswamy,
M.~C.~Gutzwiller, and G.~Casati (World Scientific, Singapore, 1991).

\bibitem{joye} Joye A., J. Stat. Phys. {\bf 75} (1994) 929.












\bibitem{kn:kamke}  Kamke E., {\it Differentialgleichungen:
L\"{o}sungsmethoden und L\"{o}sungen}, Vol. 2 (Teubner, Stuttgart,
1979).


\bibitem{Kapitaniak} Kapitaniak T., {\it Chaos for Engineers: Theory, applications, and
Control} (Springer-Verlag, Berlin, 2000).

\bibitem{Karner 1989} Karner G., Lett. Math. Phys. {\bf 17}( 1989) 329.

\bibitem{Karney1983} Karney C. F. F., Physics (Amsterdam) {\bf 8D} (1983) 360.

\bibitem{kn:kase}  Kasevich M. A., D.~S.~Weiss, and S.~Chu, Opt. Lett.
{\bf %
15} (1990) 607.

\bibitem{kn:kazan1} Kazantsev A. P., Zh. Eksp. Teor. Fiz. {\bf 66}
(1974)
1599.

\bibitem{kn:kazan}  Kazantsev A. P., G.~I.~Surdutovich, and
V.~P.~Yakovlev,
{\it Mechanical Action of Light on Atoms} (World Scientific,
Singapore,
1990).

\bibitem{khalique 2003} Khalique A., and F. Saif,
                  Phys. Lett. A {\bf 314} (2003) 37, and references there in.

\bibitem{Famlekien} Kien Fam Le, V.I. Balykin, and K. Hakuta, Phys. Rev. A {\bf 70} (2004) 063403: arXiv:quant-ph/0407107.


\bibitem{Kim 1998} Kim S. M., and J. Lee, Phys. Rev. A {\bf 61} (2000)
042102.

\bibitem{Klappauf} Klappauf B. G., W.~H.~Oskay, D.~A.~Steck, and M.~G.~Raizen,
                Phys. Rev. Lett. {\bf 81} (1998) 1203:
                 Phys. Rev. Lett. {\bf 82} (1999) 241.

\bibitem{TonVanLeewan} Koch P. M., K.~A.~H~van Leeuwen, Phys. Rep. {\bf 255}
        (1995) 289.











\bibitem{kn:lang}  Langhoff P. W., Am. J. Phys. {\bf 39} (1971) 954.

\bibitem{kn:lary} Laryushin D. V., Yu.~B.~Ovchinnikov, V.~I.~Balykin,
and V.~S.~Letokhov, Opt. Commun. {\bf 135} (1997) 138.

\bibitem{Latka 1995} Latka M., and B. J. West, Phys. Rev. Lett. {\bf
75} (1995) 4202.

\bibitem{Leibfried 1996} Leibfried D., Meekhof D.M., King B.E.,
Monroe C., Itano W.M., Wineland D.J., Phys. Rev. Lett. {\bf 77}
(1996) 4281.

\bibitem{kn:clem}  Leichtle C., I.~Sh.~Averbukh, and W.~P.~Schleich,
Phys.
Rev. Lett. {\bf 77} (1996) 3999.

\bibitem{kn:cle}  Leichtle C., I.~Sh.~Averbukh, and W.~P.~Schleich,
Phys.
Rev. A {\bf 54} (1996)a 5299.

\bibitem{kn:lilico}  Lichtenberg A. J., M.~A.~Lieberman, and
R.~H.Cohen, Physica D (Amsterdam) {\bf 1} (1980) 291.

\bibitem{kn:lieb}  Lichtenberg A. J. and M.~A.~Lieberman, {\it Regular
and
Stochastic Motion} (Springer, Berlin, 1983).

\bibitem{kn:lieb1}  Lichtenberg A. J. and M.~A.~Lieberman, {\it Regular
and
Chaotic Dynamics} (Springer, New York, 1992).

\bibitem{kn:lich1}  Lieberman M. A., and A. J. Lichtenberg,
                    Phys. Rev. A {\bf 5} (1972) 1852.

\bibitem{Lima} Lima R., and D.~L.~Shepelyansky, Phys. Rev. Lett. {\bf 67} (1991) 1377.

\bibitem{Lin 1988} Lin W. A. and L. E. Reichl, Phys. Rev. A {\bf 37} (1988) 3972.









\bibitem{Makowski 1991} Makowski A. J. and S. T. Dembi\'nski, Phys. Lett. A
                       {\bf 154} (1991) 217.

\bibitem{Mandel 1995} Mandel L., and Wolf E., Optical Coherence and
Quantum Optics (Cambridge University Press, 1995).

\bibitem{kn:Mandel}  Mandel L., Phys. Scr. {\bf T12} (1986) 34-42.

\bibitem{Marzoli 1998}  Marzoli I., F.~Saif, I.~Bialynicki-Birula,
O.~M.~Friesch, A.~E.~Kaplan, W.~P.~Schleich, Acta Physica Slovaca {\bf
48}
(1998) 323.

\bibitem{kn:mats}  The decay length of the evanescent wave is measured
upto $0.55\mu m$ in Matsudo T., H. Hori, T. Inoue, H. Iwata, Y.
Inoue, T. Sakurai,
Phys. Rev. A {\bf 55} (1997) 2406.

\bibitem{Meekhof 1996} Meekhof D. M., C. Monroe, B. E. King, W. M. Itano, and D. J. Wineland,
Phys. Rev. Lett. {\bf 76} (1996) 1796.

\bibitem{kn:mehta}  Mehta A. and J.~M.Luck, Phys. Rev. Lett. {\bf 65}
(1990)
393.

\bibitem{Meiss1985} Meiss J., and E. Ott, Phys. Rev. Lett. {\bf 55} (1985) 2741.

\bibitem{Meiss1986} Meiss J., and E. Ott, Physica D (Amsterdam) {\bf 20} (1986) 387.

\bibitem{Meystre} Meystre P., {\it Atom Optics} (Springer, Heidelberg 2001).

\bibitem{kn:mlyn} Mlynek J., V.~Balykin, and P.~Meystre, {\it Special
issue on Optics and Interferometry with Atoms}, Appl. Phys. B {\bf
54} (1992) 319.

\bibitem{Monroe 1996} Monroe C. et al, Science {\bf 272} (1996) 1131.

\bibitem{Monroe 1995} Monroe C., D. M. Meekhof, B. E. King, W. M. Itano, and D. J. Wineland,
Phys. Rev. Lett. {\bf 75} (1995) 4714.

\bibitem{Monteoliva 1998} Monteoliva D. B., B. Mirbach, and H. J. Korsch, Phys. Rev. A {\bf 57}
(1998) 746.

\bibitem{Moore1995} Moore F. L., J. C. Robinson, C. Bharucha, B. Sundaram, and
                   M. G. Raizen, Phys. Rev. Lett. {\bf 75} (1995) 4598.

\bibitem{moore} Moore F. L., J. C. Robinson, C. Bharucha, P. E. William, and
                   M. G. Raizen, Phys. Rev. Lett. {\bf 73} (1994) 2974.

\bibitem{Moskowitz} Moskowitz P. E., P.~L.~Gould, S.~R.~Atlas, and D.~E.~Pritchard,
        Phys. Rev. Lett. {\bf 51} (1983) 370.












\bibitem{ito} Ito H., T. Nakata, K. Sakaki, M. Ohtsu, K. I. Lee, and W. Jhe,
Phys. Rev. Lett. {\bf 76} (1996) 4500.

\bibitem{kn:naro}  Narozhny N. B., J.~J.~S\'{a}nchez-Mondrag\'{o}n,
and
J.~H.~Eberly, Phys. Rev. A {\bf 23} (1981) 236.











\bibitem{Oliveira 1994}  Oliveira C. R. de, I.~Guarneri, and G.~Casati,
Europhys. Lett. {\bf 27} (1994) 187.

\bibitem{kn:opat}  Opat G. I., S.~J.~Wark, A.~Cimmino, Appl. Phys. B
{\bf 54}
(1992) 396.

\bibitem{kn:ott}  Ott E., {\it Chaos in Dynamical Systems} (Cambridge
University Press, New York, 1993).

\bibitem{kn:ovchinni} Ovchinnikov Yu. B., S. V. Shul'ga, and V. I. Balykin,
J. Phys. B: At. Mol. Opt. Phys. {\bf 24} (1991) 3173.

\bibitem{kn:yub}  Ovchinnikov Yu. B., J.~S\"{o}ding, and R.~Grimm,
Pis'ma
ZETF {\bf 61} (1995) 23 [JETP Lett. {\bf 61} (1995) 21].

\bibitem{kn:ovch}  Ovchinnikov Yu., D.~V.~Laryushin, V.~I.~Balykin,
and
V.~S.~Letokhov, Pis'ma v ZETF {\bf 62} (1995) 102 [JETP Lett. {\bf 62}
(1995) 113].

\bibitem{Ovchinnikov1997}  Ovchinnikov Yu. B., I.~Manek, and R.~Grimm, Phys.
Rev. Lett. {\bf 79} (1997) 2225.









\bibitem{Paul 1990} Paul W., Rev. Mod. Phys. {\bf 62} (1990) 531.

\bibitem{kn:park}  Parker J. and C.~R.~Stroud, Jr., Phys. Rev. Lett.
{\bf 56} (1986) 716.

\bibitem{kn:pill}  Pillet P., {\it Special issue on Optics and
               Interferometry with Atoms}, J. Phys. II {\bf 4} (1994) 1877.

\bibitem{Poyatos 1996} Poyatos J.F., Walser R., Cirac J.I., Zoller
P., and Blatt R., Phys. Rev. A {\bf 53} (1996) R1966.

\bibitem{Prange 1990} Prange R. E., in {\it Quantum Chaos}, Proceedings
of the Adriatico Research Conference and Miniworkshop, Trieste,
Italy, 4 June--6 July 1990, edited by H.~A.~Cerdeira, R.~Ramaswamy,
M.~C.~Gutzwiller, and G.~Casati (World Scientific, Singapore, 1991).

\bibitem{kn:pust}  Pustyl'nikov L. D., Trudy Moskov. Mat. Ob\v {s}\v
{c}.
Tom {\bf 34}(2) (1977) 1 [Trans. Moscow Math. Soc. {\bf 2} (1978) 1].

\bibitem{pust1983} Pustilnikov L. D., Teor. Mat. Fiz.  {\bf 57} (1983) 128.

\bibitem{pust1987}  Pustilnikov L. D., Dokl. Akad. Nauk SSSR {\bf 292} (1987) 549
[Sov. Math. Dokl. {\bf 35} (1987) 88].






\bibitem{kn:mot}  Raab E. L., M.~Prentiss, A.~Cable, S.~Chu, and
D.~E.~Pritchard, Phys. Rev. Lett. {\bf 59} (1987) 2631.

\bibitem{Rasel} Rasel E. M., M.~K.~Oberthaler, H.~Batelaan, J.~Schmiedmayer,
and A.~Zeilinger, Phys. Rev. Lett. {\bf 75} (1995) 2633.

\bibitem{kn:richens}  Richens P. J. and M.~V.~Berry, Physica D {\bf 2} (1981) 495.

\bibitem{mgraizen}  Raizen M. G., Adv. At. Mol. Opt. Phys. {\bf 41} (1999) 43.

\bibitem{kn:reic}  Reichl L. E. and W.~A.~Lin, Phys. Rev. A {\bf R33},
(1986) 3598.

\bibitem{kn:reichl} Reichl L. E., {\it The Transition to Chaos in
Conservative Classical Systems: Quantum Manifestations}
                    (Springer-Verlag, Berlin, 1992).

\bibitem{renn} Renn M. J., D. Montgomery, O. Vdovin, D. Z. Anderson, C. E. Wieman,
and E. A. Cornell, Phys. Rev. Lett. {\bf 75} (1995) 3253.

\bibitem{Riedel} Riedel K., P.~T\"orm\"a, V.~Savichev, and W.~P.~Schleich,
        Phys. Rev. A {\bf 59} (1999) 797.

\bibitem{kn:hinds} Roach T. M., H.~Abele, M.~G.~Boshier,
H.~L.~Grossman,
K.~P.~Zetie, and E.~A.~Hinds, Phys. Rev. Lett. {\bf 75} (1995) 629.

\bibitem{Robinett} Robinett R. W., J. Math. Phys. {\bf 41} (2000) 1801.

\bibitem{kn:indi}  Robinson J. C., C.~Bharucha, F.~L.~Moore,
R.~Jahnke,
G.~A.~Georgakis, Q.~Niu, M.~G.~Raizen, and B.~Sundaram, Phys. Rev.
Lett.
{\bf 74} (1995) 3963.

\bibitem{kn:rosea}  Rosenbusch P., B.~V. Hall, I.~G. Hughes, C.~V. Saba,
and E.~A. Hinds, Phys. Rev. A {\bf 61} (2000) R031404.

\bibitem{kn:roseb}  Rosenbusch P., B.~V. Hall, I.~G. Hughes, C.~V. Saba,
and
E.~A. Hinds, Appl. Phys. B {\bf 70} (2000) 709.

\bibitem{Ruffo1996} Ruffo S., and D. L. Shepelyansky, Phys. Rev. Lett. {\bf 76} (1996) 3300.

\bibitem{kn:rych} Rychtarik D., B. Engeser, H.-C. Nägerl, and R. Grimm, Phys. Rev. Lett.
{\bf 92} (2004) 173003.
















\bibitem{Saif 1998}  Saif F., I.~Bialynicki-Birula, M.~Fortunato, and
W.~P. Schleich, Phys. Rev. A {\bf 58} (1998) 4779.

\bibitem{Saif 1999} Saif F., Dynamical localization and quantum revivals
in driven systems (Lehmann, Berlin 1999).

\bibitem{Saif 2000a} Saif F., K.~Riedel, B.~Mirbach, and W.~P.~Schleich,
             in Decoherence: Theoratical, Experimental and Conceptual Problems,
             ed. by Ph.~Blanchard, D.~Giulini, E.~Joos, C.~Kiefer and
             I.~-O.~Stamatescu (Springer Heidelberg 2000)a.

\bibitem{Saif 2000b} Saif F.,
Phys. Lett. A {\bf 274} (2000)b 98.

\bibitem{Saif 2000c}  Saif F., Phys. Rev. E {\bf 62} (2000)c 6308.

\bibitem{Saif 2000d} Saif F., J. Phys. Soc. Jpn. {\bf 69}8 (2000)d L2363.

\bibitem{Saif 2000e}  Saif F., G.~Alber, V.~Savichev, and W.~P.~Schleich
J. Opt. B: Quantum Semiclass. Opt. {\bf 2} (2000)e 668.

\bibitem{Saif 2001} Saif F. and A. Khalique 
                 in "International conference on Physics in Industry", ed. by
                 Anwar-ul-Haq, Mushtaq Ahmad, Karachi (2001).

\bibitem{Saif 2001a} Saif F., Fam Le Kien, and M. S. Zubairy,
                 Phys. Rev. A {\bf 64} (2001)a 043812, and references therein.

\bibitem{Saif 2002} Saif F., and M.~Fortunato, Phys. Rev. A {\bf 65} (2002) 013401.

\bibitem{Saif 2005} Saif F., J. Opt B: Quantum Semiclass. Opt. {\bf 7} (2005) S116.

\bibitem{Saif 2005a} Saif F., to be published

\bibitem{kn:sargent}  Sargent M. III, M.~O.~Scully, and W.~E.~Lamb,
{\it %
Laser Physics} (Addison-Wesley Publishing, Redwood City, 1993).

\bibitem{Scheininger} Scheininger C., and M. Kleber, Physica D {\bf 50} (1991) 391.

\bibitem{Schleich} Schleich W. P., {\it Quantum Optics in Phase Space}
        (VCH-Wiley, Weinheim 2001).

\bibitem{kn:schu}  Schuster H. G., {\it Deterministic Chaos} (VCH,
Weinheim,
1989), 2nd ed.

\bibitem{kn:scul}  Scully M. O. and M.~S.~Zubairy, {\it Quantum
Optics} (Cambridge University Press, 1997).

\bibitem{kn:saba}  Saba C. V., P.~A. Barton, M.~G. Boshier, I.~G. Hughes, P.
Rosenbusch, B.~E. Sauer and E.~A. Hinds and I.~G. Hughes, J. Phys. D {\bf
32}
(1999) R119.

\bibitem{kn:seba}  Seba P., Phys. Rev. A {\bf 41} (1990) 2306 and
references
therein.

\bibitem{kn:segev}  Segev B., R.~C\^{o}t\'{e}, and M.~G.~Raizen, Phys.
Rev.
A {\bf 56} (1997) R3350.

\bibitem{kn:seif}  Seifert W., R.~Kaiser, A.~Aspect, and J.~Mlynek,
Opt.
Commun. {\bf 111} (1994) 566.

\bibitem{kn:shore} Shore B. W.,
        {\it The theory of coherent atomic excitation},
        (New York, NY, Wiley, 1990).

\bibitem{Sigel} Sigel M. and J.~Mlynek, Physics World {\bf 6}(2) (1993) 36.

\bibitem{Sleator1} Sleator T., T.~Pfau, V.~Balykin, O.~Carnal and J.~Mlynek,
                  Phys. Rev. Lett. {\bf 68} (1992) 1996.

\bibitem{Sleator2} Sleator T., T.~Pfau, V.~Balykin, and J.~Mlynek,
                  Appl. Phys. B {\bf 54} (1992) 375.

\bibitem{kn:sodi}  S\"{o}ding J., R.~Grimm, and Yu.~B.~Ovchinnikov,
Opt.
Commun. {\bf 119} (1995) 652.

\bibitem{kn:sten}  Steane A., P.~Szriftgiser, P.~Desbiolles, and
J.~Dalibard, Phys. Rev. Lett. {\bf 74} (1995) 4972.

\bibitem{Stenholm 1992} Stenholm S., J. Mod. Opt. {\bf 39} (1992) 192.

\bibitem{Svelto} Svelto O., and D.~C.~Hanna,
            {\it Principles of lasers} 4th ed.
            (New York, NY, Plenum, 1998).

\bibitem{Szriftgiser 1996}  Szriftgiser P., D.~Gu\'{e}ry-Odelin, M.~Arndt, and
J.~Dalibard, Phys. Rev. Lett. {\bf 77} (1996) 4.













\bibitem{Tanabe2002} Tanabe S., S. Watanabe, F. Saif, and M. Matsuzawa
Phys. Rev. A {\bf 65} (2002) 033420.


\bibitem{theuer} Theuer H., R.G. Unanyan, C. Habscheid, K. Klein and K. Bergmann,
                Optics Express {\bf 4}(2) (1999) 77.

\bibitem{kn:toms} Tomsovic S. and J. Lefebvre,  Phys. Rev. Lett. {\bf 79} (1997) 3629.













\bibitem{kn:ulam}  Ulam S. M., in Proc. Fourth Berkeley Sympos. Math.
Stat.
and Probability, Vol. 3 (University of California Press, Berkeley,
1961).









\bibitem{Visscher} Visscher W. M., Phys. Rev. A {\bf 36} (1987) 5031.

\bibitem{kn:vlad}  Vladimirski\^{i} V. V., Sov. Phys. JETP {\bf 12}
(1961) 740.

\bibitem{kn:vrak}  Vrakking M. J. J., D. M. Villeneuve, and A. Stolow, Phys.
Rev. A {\bf 54} (1996) R37.








\bibitem{Wallentowitz 1995} Wallentowitz S. and W. Vogel,  Phys. Rev. Lett. {\bf 75} (1995) 2932.

\bibitem{kn:wallis}  Wallis H., J.~Dalibard, and C.~Cohen-Tannoudji,
Appl.
Phys. B {\bf 54} (1992) 407.

\bibitem{wallis 1995} Wallis H., Phys. Rep. {\bf 255} (1995) 203.

\bibitem{kn:walls} Walls D. F., and G.~J.~Milburn, {\it Quantum Optics}
                  (Springer, Heidelberg, 1994).

\bibitem{kn:whin}  Whineray S., Am. J. Phys. {\bf 60} (1992) 948.











\bibitem{kn:yan}  Yannocopoulos A. N. and G.~R.~Rowlands, J. Phys. A:
Math.
Gen. {\bf 26} (1993) 6231.

\bibitem{kn:yurke}  Yurke B., and D. Stoler, Phys. Rev. Lett. {\bf 57} (1986) 13.








\bibitem{zaslavskii} Zaslavskii G. M., and B. Chirikov, Dokl. Akad. Nauk SSSR
{\bf 159} (1964) 306
[Sov. Phys. Dokl. {\bf 9} (1965) 989].

\bibitem{zaslavsky} Zaslavsky G., Phys. Rep. {\bf 80} (1981) 157.

\bibitem{kn:Zurek} Zurek W. H.,
                  Phys. Today {\bf 44} (1991) 36;
            Progr. Theor. Phys. {\bf 89} (1991) 28.

\bibitem{Zurek 1995} Zurek W. H., and J. P. Paz, Physica D {\bf 83}
(1995) 300.


\end{thebibliography}
\end{document}